\documentclass[final,authoryear,3p,times,twocolumn]{elsarticle}

\usepackage{graphicx}
\usepackage{xcolor}
\usepackage{amssymb}
\usepackage{amsmath}

\journal{Astronomy \& Computing}

\begin{document}

\begin{frontmatter}

\title{Adventures in the microlensing cloud:\\ large datasets, eResearch tools, and GPUs}

\author{G.~Vernardos\corref{cor1}}
\ead{gvernard@astro.swin.edu.au}

\author{C.J.~Fluke}
\ead{cfluke@swin.edu.au}

\address{Centre for Astrophysics \& Supercomputing, Swinburne University of Technology, PO Box 218, Hawthorn, Victoria, 3122, Australia}

\begin{abstract}
As astronomy enters the petascale data era, astronomers are faced with new challenges relating to storage, access and management of data.
A shift from the traditional approach of combining data and analysis at the desktop to the use of remote services, pushing the computation to the data, is now underway.
In the field of cosmological gravitational microlensing, future synoptic all--sky surveys are expected to bring the number of multiply imaged quasars from the few tens that are currently known to a few thousands.
This inflow of observational data, together with computationally demanding theoretical modelling via the production of microlensing magnification maps, requires a new approach.
We present our technical solutions to supporting the GPU-Enabled, High Resolution cosmological MicroLensing parameter survey (GERLUMPH).
This extensive dataset for cosmological microlensing modelling comprises over 70,000 individual magnification maps and ${\sim}10^6$ related results.
We describe our approaches to hosting, organizing, and serving ${\sim}$30 Terabytes of data and metadata products.
We present a set of online analysis tools developed with PHP, JavaScript and WebGL to support access and analysis of GELRUMPH data in a Web browser.
We discuss our use of graphics processing units (GPUs) to accelerate data production, and we release the core of the {\tt GPU-D} direct inverse ray--shooting code (Thompson et al., 2010; Astrophysics Source Code Library, record ascl:1403.001) used to generate the magnification maps.
All of the GERLUMPH data and tools are available online from {\tt http://gerlumph.swin.edu.au}.
This project made use of gSTAR, the GPU Supercomputer for Theoretical Astrophysical Research.
\end{abstract}

\end{frontmatter}

\section{Introduction}
\label{sec:intro}
Quasar microlensing refers to the gravitational lensing effect of stellar mass objects within foreground galaxies that lie along the line of sight to multiply-imaged background quasars.
It provides a unique opportunity to study and constrain both the size and geometry of the quasar's main components \citep[see][for a review]{Schmidt2010}.
This includes investigations on scales from the broad emission--line region \citep[${\sim} 10^{17}$ cm, e.g. ][]{Sluse2012} down to the central supermassive black hole and accretion disc \citep[${\sim} 10^{14}$ cm, e.g. ][]{Dai2010}.  These physical scales correspond to typical angular scales of the order of microarcsecs, which are well below the resolution of current telescopes \citep{Rauch1991}.

There are currently ${\sim}$90 known multiply imaged quasars \citep{Mosquera2011b}, 23 of which have been studied using microlensing techniques \citep[see compilation by][]{Bate2012}.
Consequently, most investigations have focused on single objects:  the more challenging joint analysis of small collections of objects has only recently commenced \citep[e.g.][]{Morgan2010,Blackburne2011,Sluse2012,Jimenez2014}.

This situation is expected to change soon, with an anticipated increase in the number of known multiply imaged systems from a few tens to a few thousands \citep{Oguri2010}.
This is due to the commencement of  synoptic all--sky surveys, including the Pan--STARRS \citep[][]{Kaiser2002}, SkyMapper \citep[][]{Keller2007}, and Large Synoptic Survey Telescope \citep[LSST;][]{LSST2009} projects.

There is a need now to explore and understand the quasar microlensing parameter space in preparation for these future discoveries \citep[][]{Bate2012}.
The theoretical data required for this exploration, coupled with the inflow of observational data for thousands of multiply-imaged systems, will require new strategies  for effective data management to support systematic approaches to quasar modeling.
Indeed, as astronomy is now well into the petascale data era, new challenges are arising with regards to the storage, access and management of all astronomical data \citep[e.g.][]{Berriman2011}.
The traditional approach of analyzing observations, producing simulations, and comparing the two on the astronomer's desktop is now giving way to the use of remote services and resources, pushing the computation to the data.

\subsection{GERLUMPH: GPU-enabled parameter survey}
At the heart of most quasar microlensing studies lies the creation of a magnification map - a computationally demanding task, either in terms of the processing time or the system memory requirements.

A magnification map is a pixellated version of the caustic pattern in the background source plane created by the foreground microlenses, obtained using the gravitational lens equation (see Section \ref{sec:ray-shooting}).
With these maps, models of the quasar structure can be compared statistically to observations, using either the light--curve or the snapshot methods \citep[e.g. see][for some applications]{Morgan2010,Bate2008,Floyd2009}.

The majority of the available techniques for generating a magnification map \citep[e.g. ][]{Wambsganss1999,Kochanek2004,Mediavilla2011b} are based on the inverse ray--shooting technique \citep{Kayser1986}, and are single--, or multi--core central processing unit (CPU) implementations.

It was realized early on \citep[e.g.][]{Wambsganss1992} that the large--scale production of magnification maps for many multiply imaged systems would require a  computational power beyond the capabilities of the time.
In the following two decades, the focus on single-object studies meant that this limitation had minimal impact on progress in the field.
With the advent of massively-parallel, graphics processing units (GPUs), a new opportunity has arisen to accelerate the ray-shooting calculation.
This approach was demonstrated with the brute-force {\tt GPU-D} code by \citet{Thompson2010}, which was later compared with the \cite{Wambsganss1990a,Wambsganss1999} single--core tree-code by \citet{Bate2010}.

The aim of the GPU Enabled, High Resolution, cosmological MicroLensing parameter survey (GERLUMPH), is to provide a theoretical resource, consisting of tens of thousands of magnification maps, to use in preparing for the synoptic survey era of microlensing.
This open data resource is complemented by online analysis tools supporting modeling of the known and discovered microlensed systems.
GERLUMPH acts as a moderate--data size (${\sim} 30$ Terabytes) case study, where sharing of data was a guiding principle.
All of the GERLUMPH data products and online analysis tools are freely and publicly accessible from:
\begin{center}
{\tt http://gerlumph.swin.edu.au}
\end{center}

Compared to a more general simulation-- or theory--based virtual observatory, which might need to cater for a very wide range of analysis tasks, there is only a limited set of standard analysis tasks that are used by the microlensing community.
This made it much more practical to build in these tools and provide them to the user through a web browser.
Development of the browser--based solution was commensurate with the first stable release of the WebGL\footnote{Web Graphics Library: {\tt www.khronos.org/webgl}} JavaScript application programming interface, so we took advantage of this to investigate rich, interactive visualisation tools for compatible browsers.

\subsection{GPU Supercomputing}
While access to a single GPU can provide significant speed-ups to existing CPU-based solutions, access to a computing cluster equipped with GPUs provides ${\mathcal{O}}(100)$ Tflop/s performance at the fraction of the cost of the equivalent CPU system.

Throughout this work, we have used the GPU--Supercomputer for Theoretical Astrophysics Research (gSTAR), located at Swinburne University of Technology.
The gSTAR facility comprises 53 CPU--core nodes,  50 of which are equipped with two NVIDIA C2070 GPUs; the remaining 3 nodes comprise 7 NVIDIA M2090 GPUs each.
Additionally, gSTAR is connected to a ${\sim}$1 Petabyte storage system, using the Lustre\footnote{\tt http://www.opensfs.org/lustre/} parallel filesystem.

75 per cent of the computing time on gSTAR is available on a competitive basis, with requests for computing time governed by the Astronomy Supercomputing Time Assignment Committee (ASTAC).


\subsection{Overview}
In this work, we describe the data management infrastructure and remote analysis services developed for GERLUMPH.
The scientific motivation and outcomes of GERLUMPH are described elsewhere \citep{Bate2012,Vernardos2013,Vernardos2014a}.

In Section 2 we describe the {\tt GPU-D} inverse ray--shooting technique and present updated benchmarks, while the core GPU code is presented in detail in \ref{app:release} and released to the community for examination and further enhancement.
Our approach to data management and storage is described in Section 3 and \ref{app:compression}.
The GERLUMPH online eResearch tools are presented in Section 4.
Discussion and conclusions follow in Sections 5 and 6.

\section{GPU-accelerated microlensing}
How should one adapt or develop code for a GPU to accelerate a scientific computation when the existing best software solution is designed for either a single or low number of CPU compute cores?
\cite{Barsdell2010} advocate the use of an algorithm analysis strategy, whereby alternative algorithms are chosen that more closely match the massively parallel GPU architecture.
A compelling class of alternative algorithms are brute force or direct calculation solutions, often related to the way a particular scientific computation was originally proposed - before highly optimised or approximate solutions were investigated.
In certain cases, brute force algorithms present a simplified coding option for GPUs, providing sufficient acceleration to solve problems that were not feasible with single-core CPU-only solutions \citep{Fluke2011}.

\subsection{Brute force ray-shooting}
\label{sec:ray-shooting}
\begin{figure}
\begin{center}
\includegraphics[width=7.6cm, angle=0]{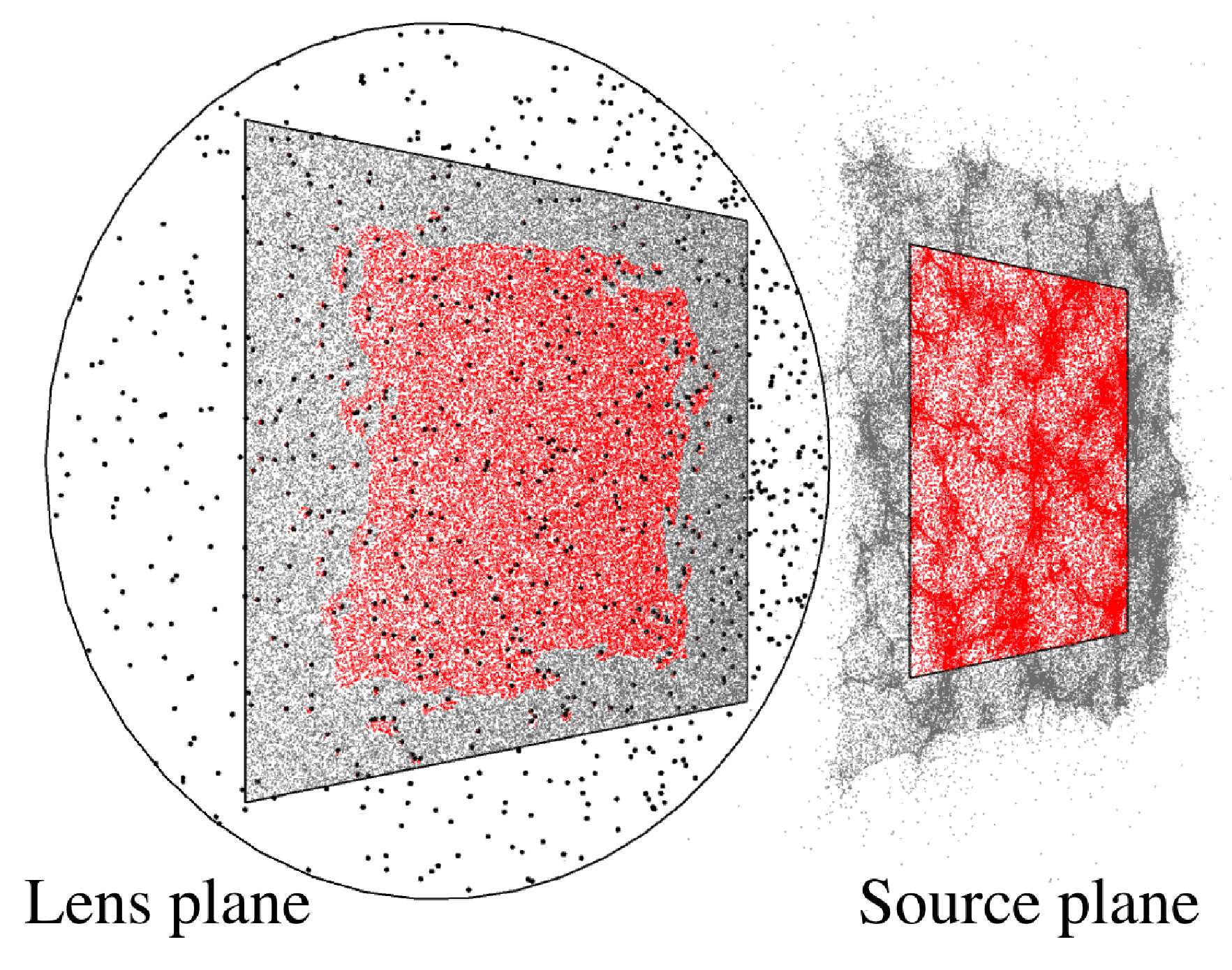}
\caption{Schematic representation of ray--shooting. Microlenses (black dots) are distributed in a circle of area $A$ on the lens plane (see equation \ref{eq:Nl}). A grid of light-rays (grey and red dots) is projected backwards from the observer through the lens plane, where the deflection by each lens is calculated for each ray using equation (\ref{eq:lenseq}), and mapped on to the source plane. A rectangular area on the source plane is selected (red points) away from edge effects, divided into pixels, and used to calculate the magnification from equation (\ref{eq:mu}).}
\label{fig:schematic}
\end{center}
\end{figure}

One such brute force algorithm was described and tested in \cite{Thompson2010}: inverse ray-shooting for gravitational microlensing \citep[see][for early implementations]{Kayser1986}.
Here, large numbers (${\sim} 10^{9}$) of light rays are projected from the observer, through the lens plane, where they are each deflected by $N_{*}$ individual lenses according to the gravitational lens equation, and accumulated on a gridded source plane.
For the specific case of cosmological microlensing by $N_*$ compact, point-mass objects in the presence of a smooth matter distribution, and an external shear, ${\gamma}$, the gravitational lens equation is:
\begin{equation}
\label{eq:lenseq}
\boldsymbol{y} = 
\begin{pmatrix}
1-\gamma & 0\\ 
0 & 1+\gamma 
\end{pmatrix}
\boldsymbol{x} -
\kappa _{\textrm{s}} \boldsymbol{x} - 
\sum_{i=1}^{N_*}m_{i} \frac{\left ( \boldsymbol{x - x _{i}} \right )}{\left | \boldsymbol{x -\boldsymbol{x _{i}}} \right | ^{2}} \, ,
\end{equation}
This equation relates the position of a light ray in the source plane, ${\mathbf{y}}$, to a lens plane location, ${\mathbf{x}}$.
The total convergence, ${\kappa} = {\kappa_{s}} + {\kappa_{*}}$, has contributions from both compact objects, ${\kappa_{*}}$, and smooth matter, ${\kappa_{s}}$.
It is convenient to introduce the smooth matter fraction:
\begin{equation}
s = \frac{\kappa_{\rm s}}{\kappa},
\end{equation}
allowing the definition of a map in terms of the three parameters ${\kappa},{\gamma},s$.
The number of microlenses, $N_*$, is:
\begin{equation}
\label{eq:Nl}
N_* = \frac{\kappa_* A}{\pi \langle M \rangle} \, ,
\end{equation}
where ${\langle} M {\rangle}$ is the mean mass of the microlenses, and $A$ is the area where they are distributed.
A schematic representation of the direct inverse ray-shooting technique can be seen in Figure \ref{fig:schematic}.

The number of light rays reaching each pixel of the source plane, $N_{ij}$, compared to the number of rays that would have reached each pixel in the absence of microlensing, $N_{\rm avg}$, gives the local magnification value, ${\mu_{ij}}$:
\begin{equation}
\mu_{ij} = \frac{N_{ij}}{N_{\rm avg}}.
\label{eq:mu}
\end{equation}
A magnification map, or just map, is a pixellated version of the source plane with fixed width and resolution.
The width of the map is measured in units of the Einstein radius, $R_{\rm Ein}$, the radius of the symmetric ring that occurs when a source is directly aligned with a gravitational lens or microlens:
\begin{equation}
\label{eq:rein}
R_{\rm Ein} = \sqrt{ \frac{D_{\rm os}D_{\rm ls}}{D_{\rm ol}} \frac{4G\langle M \rangle}{c^2} } \, .
\end{equation}
This quantity depends on the angular diameter distances from observer to lens, $D_{\rm ol}$, observer to source, $D_{\rm os}$, and lens to source, $D_{\rm ls}$, and the mean mass of the microlenses $\langle M \rangle$.
Once a value is set for $R_{\rm Ein}$, each map pixel will correspond to a physical length e.g. the pixels of a 25-$R_{\rm Ein}$ wide map with a 10000-pixel resolution would span 1.25${\times}10^{14}$ cm, for a typical value of $R_{\rm Ein} = 5 {\times} 10^{16}$ cm (see Section \ref{sec:data} and Table \ref{tab:components} for an explanation of these values).

The deflection calculation in equation (\ref{eq:lenseq}) requires ${\approx} 10$ floating point operations for each lens, and for simulations of realistic microlensing systems, $N_{*} {\sim} 10^3 - 10^5$ is typical.
As each light ray deflection is independent of all other light rays,  the ray-shooting algorithm is ``embarassingly parallel'' \citep{Barsdell2010} and thus highly suitable for implementation on a GPU.

\subsection{Brute force benchmarks}
The \cite{Thompson2010} brute force ray-shooting code, {\tt GPU-D}, has now been used to generate more than 70,000 magnification maps.
The validity of the code was established by \cite{Bate2010} through direct comparison with results from the \cite{Wambsganss1990a,Wambsganss1999} tree-code.
Additional testing has been performed in \cite{Vernardos2013}, investigating the dependence of magnification probability distributions on the random positions of individual microlenses, and in \cite{Vernardos2014a}, comparing magnification probability distributions for maps that are considered equivalent according to the coordinate transformation imposed by the mass-sheet degeneracy.
We present details on the released version of the {\tt GPU-D} code in \ref{app:release}, encouraging others to use, improve and further test our solution.

\cite{Thompson2010} found that a brute force implementation of microlensing ray-tracing gave a speed-up of 125 times relative to a single-core CPU implementation.
However, this is a highly unfair comparison, as the single-core brute force solution is known to be computationally inefficient.
A better benchmark is to compare this solution to the highly optimised, yet still single-core CPU-only, tree-code of \cite{Wambsganss1990a,Wambsganss1999} which has been used extensively in astronomy for more than two decades.
This latter code uses a single-core CPU solution, and so has not been able to take advantage of incremental hardware speed-ups: clock speeds for single CPU cores plateaued around 2004.
An alternative MPI-based tree-code solution for a distributed computing cluster was implemented by \cite{Garsden2010}.

\cite{Bate2010} showed that for a reasonable portion of the parameter space of interest to cosmological microlensing, the brute force solution was faster than the tree-code.
This was based on tests using an NVIDIA S1070 Tesla unit, released in 2008.
\cite{Bate2010} predicted  that by using newer generations of GPUs, there would be an immediate additional speed-up.
We now present up-to-date benchmarks of the {\tt GPU-D} code, comparing the performance on 5 generations of NVIDIA GPUs (see Table \ref{tab:cards} for details).
For graphics cards with multiple GPUs, we only use one GPU for benchmarking.

\begin{figure}
\begin{center}
\includegraphics[scale=0.185]{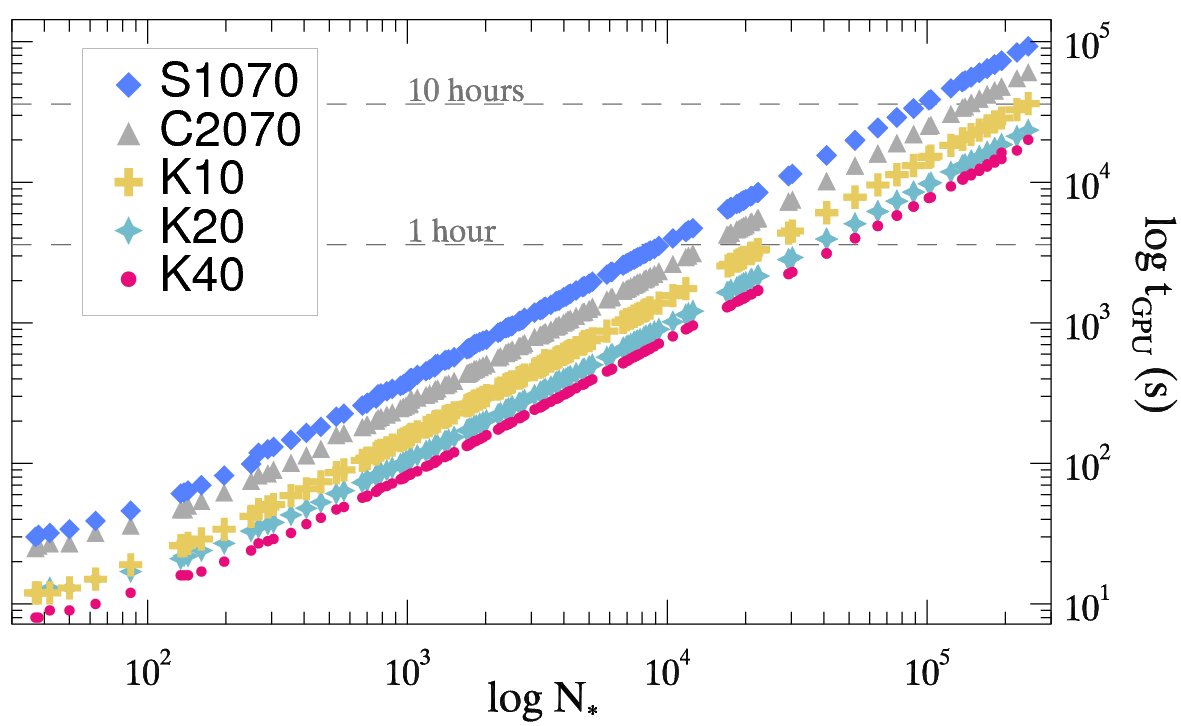}
\caption{The computational time scales linearly with the number of microlenses. The same set of 170 magnification maps was produced on 5 generations of GPUs, with an increasing number of threads (see Table \ref{tab:cards}). Our computations saturate the GPU devices for $N_* {\approx} 300$. Newer cards with more threads are faster.}
\label{fig:speedup}
\end{center}
\end{figure}

The key feature of each card is the number of threads, which is the number of instances of a GPU-code that can be executed in parallel.
To test the speed-up provided by different GPU cards, we used a set of 170 magnification map simulations with varying numbers of microlenses as a benchmark.
Increasing the number of microlenses (equation \ref{eq:Nl}) for which we have to directly solve equation (\ref{eq:lenseq}), will linearly increase the computational time.
This linear increase can be seen in Figure \ref{fig:speedup}, where we show the computational time as a function of the number of microlenses.

As the GPU cards become more powerful, the {\tt GPU-D} code has executed faster: our benchmark problem took 20 days to complete on a 2008 S1070 card, while it took 4.9 days on a 2013 K40 card, which has more than ten times the number of threads (see Table \ref{tab:cards}).
This constitutes a speed-up factor of 4 gained in roughly 5 years.
We point out here that we used the same code on each card, without any modifications to take advantage of new hardware or software features.

\begin{table}
\begin{center}
\caption{Five generations of NVIDIA GPUs used to benchmark the {\tt GPU--D} brute force ray--shooting code. The main feature of the cards that increases the peak computational power, measured in Gigaflop/s, is the number of Threads. In the last column, is the total time to generate a set of 170 benchmark maps.}
\label{tab:cards}
\begin{tabular}{lllll}
Card 			& Year 	& Peak 	& Threads & Time (days) \\
\hline
S1070$^{\dag}$	& 2008	& 1.04	& 240	  & 20   \\
C2070 			& 2010	& 1.29 	& 448	  & 13.8 \\
K10$^{\dag}$	& 2012	& 2.67 	& 1536	  & 10.6 \\
K20 			& 2012	& 4.10 	& 2496	  & 5.9  \\
K40 			& 2013	& 5.36	& 2880	  & 4.9  \\
\hline
\end{tabular}
$^{\dag}$the S1070 and K10 cards contain more than one GPU; the results reported here are only for one GPU.
\end{center}
\end{table}

\subsection{Experiences using gSTAR}

\begin{figure}
\begin{center}
\includegraphics[scale=0.18]{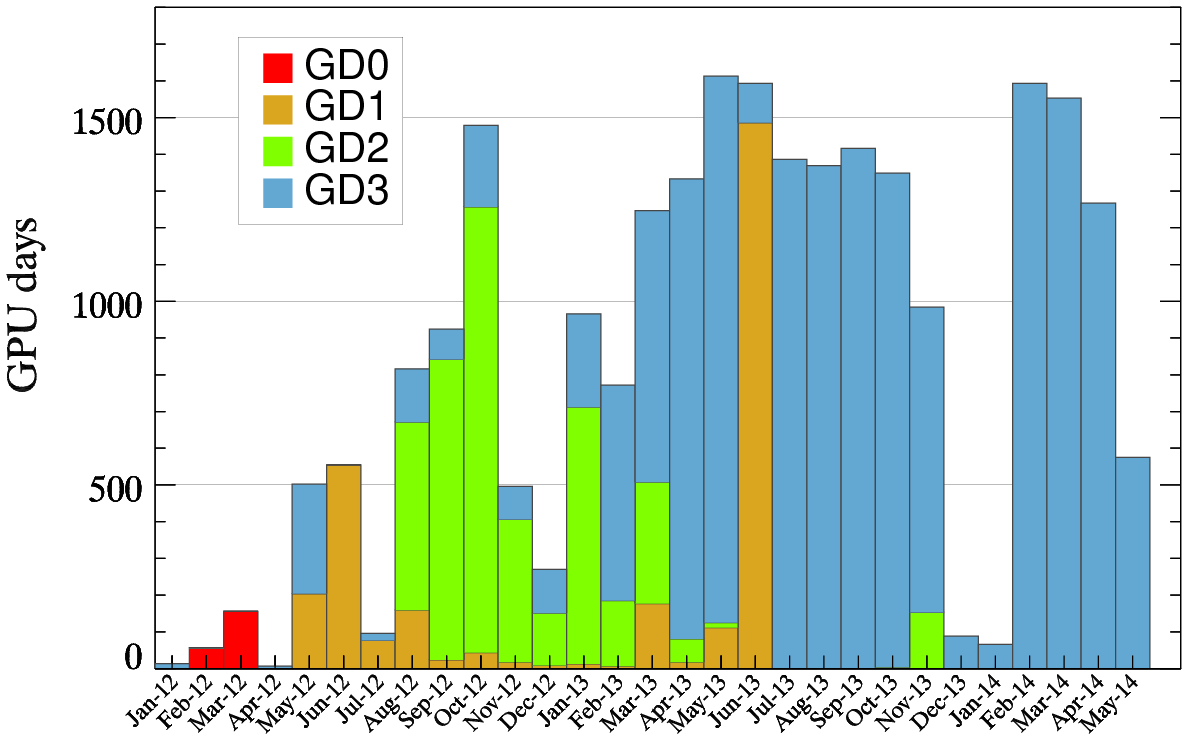}
\caption{Scheduling of the GERLUMPH computations on gSTAR. gSTAR was officially opened on the 1st of May 2012. There were two major downtimes between November and December 2012, and December 2013 and January 2014. An up-to-date version of this graph can be found at {\tt http://gerlumph.swin.edu.au/status/}, as more calculations are being completed.}
\label{fig:timeused}
\end{center}
\end{figure}

In Figure \ref{fig:timeused} we see how the GERLUMPH datasets have used the computational time available on gSTAR: GD0 was generated during the preliminary phase of gSTAR, before its official opening on May 1st 2012, GD1 and GD2 were mostly completed within the following year, and GD3 will be completed by mid 2014 (see Section \ref{sec:data} and Table \ref{tab:datasets} for more details on the GERLUMPH datasets).
We have used the equivalent of 21,000 CPU days so far, ${\sim}$9,000 of which have been allocated to GERLUMPH via ASTAC.
Sharing the gSTAR nodes with other users lead to identifying issues with the GPU device allocation, which have been resolved by working closely with the supercomputer support team.

We can compare our timing results for conducting a parameter survey with the strategies proposed by \cite{Bate2012}.
The general idea is to start by producing maps at the positions of the currently known macromodels and then move progressively outwards to cover the rest of the parameter space (see Figure \ref{fig:coverage}).
The authors suggested different stages of the survey, depending on the parameters examined and the expected specifications of the gSTAR supercomputer.
Their {\it Stage 2} includes all maps located at and around the known macromodel positions, with ${\Delta}{\kappa}={\Delta}{\gamma}=0.01$.
Using the timing relation derived by \cite[][equation 7]{Thompson2010} for a S1070 card, they estimated a computational time of 177 days, using a computational power of 100 teraflops/s.
We carried out these calculations in 74 days using the newer C2070 cards on gSTAR, at 64 Teraflop/s (maximum computational power allowed per user). This result allowed us to design and proceed with the exploration of other areas of parameter space, presented in detail in Section \ref{sec:data}.


\subsection{Convolutions}
\label{sec:convolutions}
A further opportunity to utilize GPUs in quasar microlensing exists.
If the physical size of the accretion disc is less than the physical size of a map pixel, then we are dealing with a point-source and we can use the magnification map as is to extract observable quantities without further processing.
Otherwise, we are dealing with a finite-size source, and the map must be convolved with the physical source profile before we can proceed.

A convolution between a source profile and a map, for a given value of $R_{\rm Ein}$, can be performed once the source profile is transformed into a convolution kernel.
To achieve this, the projected two-dimensional source profile is binned in pixels of equal physical size to the map pixels, and then appended with empty (zero) pixels to match the pixel dimensions of the map.
Now, the convolution kernel and the map are images (pixel arrays) of the same size and the convolution theorem can be applied:
\begin{equation}
\label{eq:conv}
C = M \ast K = F^{-1} [ F(M) F(K) ]
\end{equation}
where C is the convolved map, M is the original map, K is the convolution kernel, and $F$ and $F^{-1}$ are the forward and inverse Fourier transforms respectively.

Although generating a high-resolution magnification map typically takes from a few hours to a few days, depending on the number of microlenses, convolution with a source profile takes a few seconds on the GPU.
The Fast Fourier Transform algorithm \citep[FFT,][]{Press1992} is a widely used algorithm that has already been implemented for GPUs.
Using the cuFFT\footnote{\tt http://docs.nvidia.com/cuda/cufft/} library, which includes a tested and well-documented CUDA implementation of the FFT, we perform a convolution for $10000^2$-pixel maps in ${\sim}$3s.
For comparison, the same convolution using the the CPU version of FFT, takes about a minute.

\subsection{Parameter space exploration}
On the theoretical side, quasar microlensing simulations are intrinsically complex, with three major physical systems having to be modelled simultaneously i.e. the accretion disc, the galaxy-lens (macrolens) and the microlenses.
A morphological analysis approach would be well-suited for this problem: identify the parameter combinations that lead to possible solutions and eliminate those which do not, rather than simplifying the problem by reducing the number of parameters \citep[see][]{Zwicky1969,Ritchey2006}.
Therefore, an exploration of the microlensing parameter space, the most computationally demanding stage of the process, is warranted.

The parameters involved in generating a microlensing magnification map can be divided into three categories:
\begin{enumerate}
\item {\bf Macromodel (external) parameters.} These are the convergence and shear, ${\kappa}$ and ${\gamma}$, coming from the macromodels of the galaxy-lens, together with the smooth matter fraction, $s$, that separates ${\kappa}$ into compact and smooth matter components (see Section \ref{sec:ray-shooting}).
\item {\bf Parameters of the microlenses}, namely, the mass-spectrum and positions of the microlenses.
\item {\bf Map characteristics}, including the map width, pixel resolution and average number of light rays, $N_{\rm avg}$. Wide maps (24$R_{\rm Ein}$ or more) allow for investigation of large sources ($\geq R_{\rm Ein}$) and high resolution reveals details of the background source \citep[see the relevant discussion in][]{Bate2012}. Finally, the more rays shot, the better the statistical accuracy of the map.
\end{enumerate}

Performing such explorations in the past was time consuming \citep{Wambsganss1992}, and the only practical option was single object modelling, in line with the observations of the time that also focused on single objects.
However, with the advent of GPUs and other computational accelerators it is now possible to undertake a detailed systematic investigation of the parameter space, in preparation for the imminent era of new discoveries.

In Table \ref{tab:studies}, we show an example of the transition in the total number of maps produced, from single object and limited parameter space studies to extended parameter space explorations, as GPU computational power became available.
A GPU-based high performance computing cluster makes it possible to scale up the number of simulations, allowing parameter space explorations that would have taken years, or decades, to be completed in timescales of weeks or months \citep[see][]{Bate2012}.

\begin{table*}
\begin{center}
\caption{Number of maps used in parameter space and single object studies, compared to the total of GERLUMPH maps.}
\label{tab:studies}
\begin{tabular}{rlll}
Publication					&	Number of maps	&	Technology	&	Approach	\\
\hline
\cite{Wambsganss1992}		&	64				&	CPU			&	parameter space	\\
\cite{Lewis1995}			&	61				&	CPU			&	parameter space	\\
\cite{Mediavilla2009}		&	580				&	CPU			&	sample of 20 objects \\
\cite{Poindexter2010a}		&	1320			&	CPU			&	single object 	\\
\cite{Vernardos2013}		&	2550			&	GPU			&	parameter space	\\
\cite{Vernardos2014a}		&	12342			&	GPU			&	parameter space	\\
\hline
GERLUMPH					&	$>70,000$		&	GPU			&	parameter space	\\
\end{tabular}
\end{center}
\end{table*}

\section{Data management}
The GERLUMPH data consist of magnification maps and convolution results.
The total information in the GERLUMPH maps is of the order of 7 Terapixels, spread across ${\sim}$70,000 maps, requiring ${\sim}$25 Terabytes of uncompressed storage space.
An additional ${\sim}$0.5 Terabytes of data is required for support files, such as low-resolution preview icons, probability distributions, log files, etc.
The number of convolutions between maps and accretion disc profiles for different values of $R_{\rm Ein}$ is of the order of $10^6$.
The convolution output is directly processed and a number of quantities of interest are stored e.g. probability distributions, flux locations, etc, which amount to a final data size of 5 Megabytes per convolution, on average.

The amount of data (Section \ref{sec:data}), and the specific dependence on combinations of map, disc profile and convolution parameters, raised the need for a suitable data storage, management and access scheme.
This is achieved by managing meta-data in a MySQL relational database (Section \ref{sec:databases}), and considering suitable data formats and methods of compression (\ref{app:compression}).

\subsection{The GERLUMPH datasets}
\label{sec:data}
\begin{figure}[t]
\begin{center}
\includegraphics[width=0.47\textwidth]{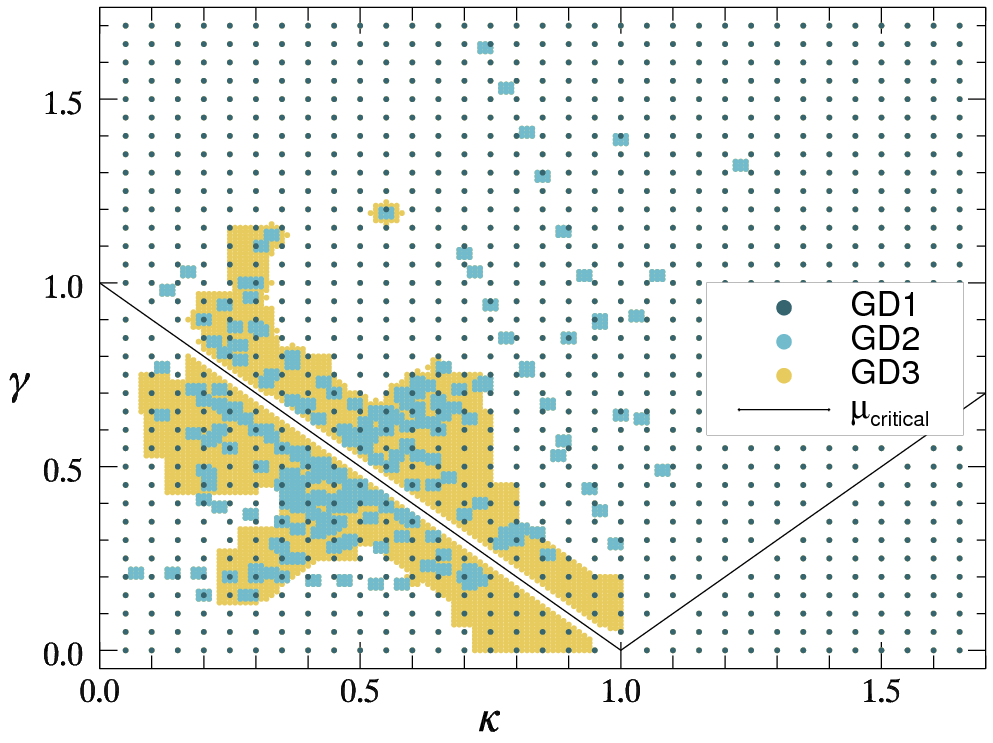}
\caption{Microlensing parameter space covered by the GERLUMPH datasets. GD1 maps (dark blue) are distributed uniformly (${\Delta} {\kappa} = {\Delta} {\gamma} = 0.05$), GD2 maps (light blue) are found at, and around, the locations of macromodels from the literature \citep{Bate2012}, with small perturbations (${\Delta} {\kappa} = {\Delta} {\gamma} = 0.01$), and GD3 maps (yellow) densely cover extended regions of parameter space (with ${\Delta} {\kappa} = {\Delta} {\gamma} = 0.01$). The black line is the critical line where ${\mu_{\rm th}} {\rightarrow} {\infty}$, with ${\mu_{\rm th}}$ being the macro-magnification. An up-to-date version of this graph can be found at {\tt http://gerlumph.swin.edu.au/status/}, as more of parameter space keeps being covered.}
\label{fig:coverage}
\end{center}
\end{figure}

The GERLUMPH magnification maps are divided into four distinct datasets, summarized in Table \ref{tab:datasets} and plotted in the ${\kappa},{\gamma}$ parameter space in Figure \ref{fig:coverage}.
Each GERLUMPH dataset is targeted at answering a specific scientific question relevant to cosmological microlensing:
\begin{enumerate} 
\item GD0 constitutes a preliminary phase of GERLUMPH at a lower resolution ($4096^2$ pixels per map). Its main goal is to investigate the effect of varying the microlens positions of the maps via the first ever uniform exploration of the ${\kappa},{\gamma}$ parameter space \citep{Vernardos2013}. GD0 was used to test practical issues related to data production and management, together with supporting gSTAR's early stages of deployment and debugging.
\item GD1 provides the first extensive coverage of the microlensing parameter space at high resolution ($10000^2$ pixels per map). The science goal of GD1 is to investigate the effect of systematically including smooth matter in the ${\kappa}, {\gamma}$ parameter space \citep{Vernardos2014a}.
\item GD2 targets 23 known multiply imaged systems and their corresponding 275 published macromodel ${\kappa}, {\gamma}$ values \citep{Bate2012}. Its main goal is to investigate the effect of small perturbations (${\Delta}{\kappa}={\Delta}{\gamma}=0.01$) on map properties, and subsequently on accretion disc properties sensitive to microlensing.
\item GD3 bridges the gap between GD1 and GD2, creating large areas of parameter space densely covered by maps (${\Delta}{\kappa}={\Delta}{\gamma}=0.01$), and investigating how well constrained macromodel ${\kappa},{\gamma}$ values have to be in large areas of the parameter space.
\end{enumerate}
Besides their intended scientific goals, subsets of GERLUMPH maps can be used for detailed studies of particular microlensed systems, or for theoretical studies across the parameter space.

\begin{table*}
\caption{The GERLUMPH magnification map datasets.}
\begin{center}
\begin{tabular}{ r l l l l l }
											&	GD0			&	GD1 		&	GD2				&	GD3$^*$			&	Total	\\
\hline
number of maps								&	2550		&	12342		&	18271			&	${\sim}$37000	&	70000	\\
GPU days									&	212			&	2903		&	4722			&	${\sim}$16900	&	12110	\\
data size (TB)								&	0.17		&	4.7			&	7				&	${\sim}$13.5	&	25		\\
unique ${\kappa},{\gamma}$ combinations 	&	170			&	1122		&	1661			&	${\sim}$3255	&	-		\\
${\kappa},{\gamma}$ coverage				& 	uniform 	& 	uniform		&	specific systems&	bridging the gap&	-		\\
$s$											&	1			&	11			&	11				&	11				&	-		\\
resolution									&	$4096^2$	&	$10000^2$	&	$10000^2$		&	$10000^2$		&	-		\\
width $(R_{\rm Ein})$						&	24			&	25			&	25				&	25				&	-		\\
realizations								&	15			&	1			&	1				&	1				&	-		\\
\hline
\end{tabular}
$^{*}$in progress.
\end{center}
\label{tab:datasets}
\end{table*}

For studying finite sized sources the choice of the three major components required to perform a convolution is described below and is summarized in Table \ref{tab:components}:
\begin{enumerate} 
\item
Any map, or set of maps, from the GERLUMPH datasets can be used, depending on the aspect of accretion disc modelling under investigation e.g. single system or parameter space, low or high resolution, effect of smooth matter, systematics of lens positions, etc.
\item
To determine the $R_{\rm Ein}$ of a gravitationally lensed system, measurements of both the source and lens redshifts are required.
The CASTLES\footnote{\tt http://www.cfa.harvard.edu/castles/} survey \citep{Falco2001} consists of 59 systems with both required redshifts measured, while \cite{Mosquera2011b} have used estimates of lens redshifts to construct a sample of 87 lensed quasars.
The mean value and standard deviation of $R_{\rm Ein}$ from the two samples is 5.35${\pm}$2.20${\times}$10$^{16}$ cm and 5.11${\pm}$1.88${\times}$10$^{16}$ cm respectively (for ${\langle} M {\rangle} = 1$ M$_{\odot}$ and $H_{\rm 0} = 70$ km s$^{-1}$ Mpc$^{-1}$, see equation \ref{eq:rein}).
We can therefore say that a value of $R_{\rm Ein}{\sim}$5${\times}$10$^{16}$ cm is typical \citep[see also][]{Bate2012}, with the majority of systems lying between 1${\times}$10$^{16}$ and 9${\times}$10$^{16}$ cm. 
This motivates the range of values we have explored; a prominent exception is Q2237+0305 with $R_{\rm Ein}=$18.1${\times}$10$^{16}$ cm, which is unusually high because the galaxy-lens happens to be very close to us at redshift $z=0.04$.
\item
Realistic state--of--the--art accretion disc models may include warped discs, spinning black holes, relativistic effects, etc \citep[for a review see][]{Abramowicz2013}.
Currently, it has not been demonstrated that microlensing is sensitive enough to detect the observational signatures of such features \citep[e.g.][]{Mortonson2005}.
Instead, using simple geometrical disc profiles greatly reduces the size of the parameter space that needs to be investigated, while still spanning a wide range of possibilities.
Example profiles include a uniform disc, a Gaussian disc, a power-law \citep[the thin-disc model,][]{Shakura1973}, or profiles with a central brightness depression corresponding to the event horizon of the supermassive black hole.
The size of the profile, a more important parameter than the shape, is allowed to vary between 5${\times}$10$^{14}$ to 5${\times}$10$^{16}$ cm (by size we mean the radius at which the brightness essentially drops to zero, as opposed to the half-light radius).
This range allows us to study physical scales from the X-ray emitting region of the disc, which is expected to be small and centrally located \citep[${\sim} 10^{14}$ cm, e.g.][]{Dai2010}, out to the broad emission-line region \citep[${\sim} 10^{17}$ cm, e.g.][]{Sluse2012}.
\end{enumerate}

Ideally, one would keep the result of each convolution i.e. a convolved map, and use this for future extraction of arbitrary properties that can be compared to observations.
However, keeping each convolved map will quickly lead to data sizes beyond our storage capabilities; $10^6$ convolved maps, each 381 Megabytes in size, require ${\sim}$4 Petabytes of final data size to store (c.f. gSTAR has total storage disk space of ${\sim}$1 Petabyte).
Therefore, each convolved map has to be processed directly and then discarded.
Given the speed at which convolutions can be calculated on the GPU (${\sim}$3 s/map), this is a reasonable strategy.

For each convolved map, we calculate and store the magnification probability distribution (see Section \ref{sec:maptable}).
Additionally, we produce and store flux ratio data viz. the location and magnification value of typically a few thousand pixels from the convolved map, to allow for accretion disc studies of the type performed by \cite{Bate2008}, or \cite{Floyd2009}.
At this stage we choose not to generate any light-curve data from the convolved maps, as this further increases the storage requirements and is more closely linked to customized analysis fitting processes for studying specific systems \citep[e.g., see][]{Kochanek2004}.  We note in passing that there are opportunities for additional GPU-acceleration of the light-curve analysis and fitting process in the future.

\begin{table}
\caption{Summary of the GERLUMPH convolution components. Size denotes the radius where the disc brightness profile drops to zero. The shape of the profile can vary from a simple geometrical to a more realistic configuration.}
\begin{center}
\begin{tabular}{ r l }
\hline
maps:				&	${\sim}$70,000 GERLUMPH maps 		\\
Einstein radii:		&	1 $<R_{\rm Ein}<$ 9  [$10^{16}$ cm] 	\\
disc sizes:			&	0.5 $<$ size $<$ 50  [$10^{15}$ cm] 	\\
disc shapes:		&	uniform disc, gaussian, etc. 		\\
\hline
\end{tabular}
\end{center}
\label{tab:components}
\end{table}

\subsection{The GERLUMPH databases}
\label{sec:databases}
The GERLUMPH data are organized in a MySQL\footnote{\tt http://www.mysql.com/products/community/} relational database.
MySQL is a popular, open-source, well-documented, database technology and data storage system, which has been used in astronomy before \citep[e.g. ][]{Lemson2006}.
The database tables hold the metadata for each map, or convolution, together with the corresponding index of the output directory.

\begin{figure}[t]
\begin{center}
\includegraphics[scale=0.25]{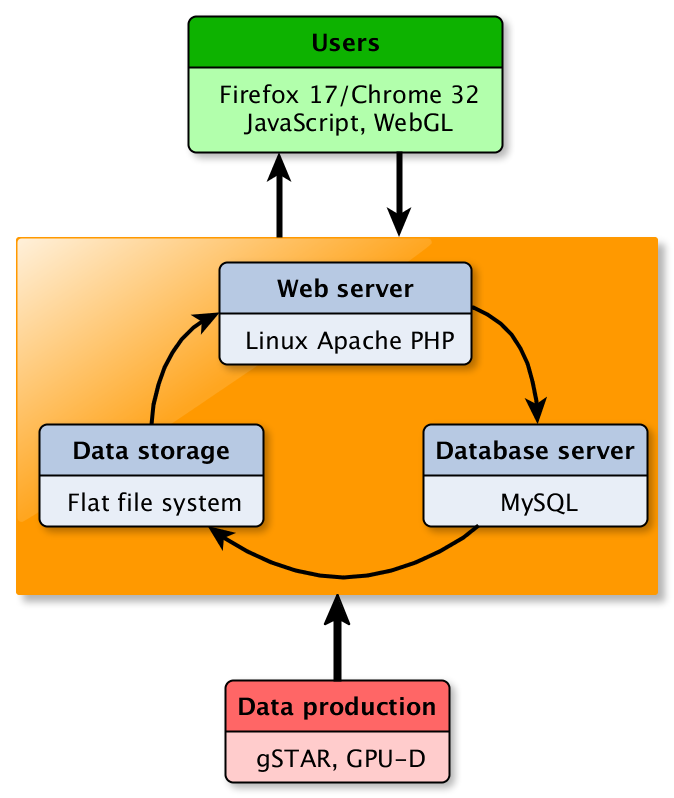}
\caption{A schematic representation of the flow of data from production to the user. The GERLUMPH implementation can be seen under each component. Data are produced on a supercomputer, and are subsequently moved to the storage devices, with the metadata put into a database. The web server, database server, and stored data can physically lie on the same, or different machines. The user makes a request at the server, which contacts the database to get the metadata, and the storage devices to retrieve the actual data, before returning the result to the user.}
\label{fig:flow}
\end{center}
\end{figure}

The actual data do not reside in the database itself.
Instead, they are located in indexed directories, which keeps the resulting disk size of the database small (a few Megabytes).
The physical location of the data is on gSTAR's storage disks, under a flat file system: all indexed output directories are stored under a single directory for magnification maps, convolutions, etc.
This approach allows us the flexibility of being able to move the data to appropriate storage devices, while the database and web servers (see Section \ref{sec:eTools}) can run on separate machines.
A schematic flow of data from their production on gSTAR to the user can be seen in Figure \ref{fig:flow}.

\begin{figure*}[t]
\begin{center}
\includegraphics[scale=0.25]{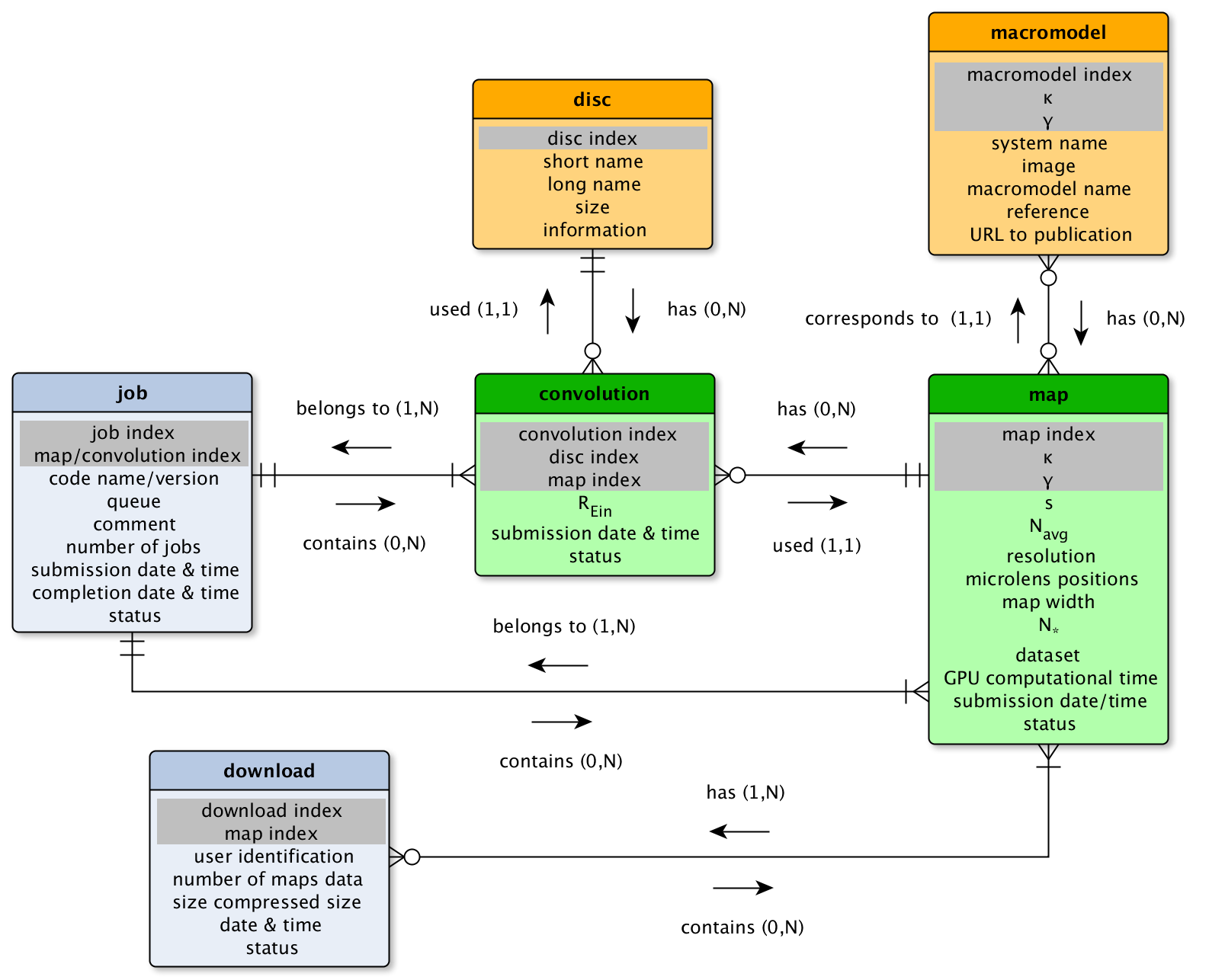}
\caption{The entity-relationship diagram for the GERLUMPH database. Entities (tables) are connected to each other by relations, such as {\it corresponds to} or {\it has}. Relations can have degrees, e.g. one-to-one (1,1), one-to-many (1,N), or none-to-many (0,N). We have used the Crow's foot and the (min,max)-notation to denote relationships on the diagram. The primary keys and common attributes between entities are shown in grey.}
\label{fig:ER}
\end{center}
\end{figure*}

In the following, we describe the GERLUMPH relational database in terms of the entity-relationship model \citep{Chen1976,Barker1990}.
It consists of six entity types:
\begin{enumerate}
\item `map' is a magnification map.
\item `disc' is an accretion disc profile.
\item `convolution' is a convolution between a map and a disc profile.
\item `macromodel' is the ${\kappa}, {\gamma}$ values at the position of an image of a multiply imaged quasar, derived from macromodels of the lens galaxy.
\item `job' is a batch of map or convolution simulations submitted to the supercomputer.
\item `download' is one or more maps selected for download by a GERLUMPH user.
\end{enumerate}
The `map' and `convolution' types constitute our actual results, the `disc' and `macromodel' types are complementing our main data, while the `job' and `download' types are for management purposes only.
Each entity type is represented by a table whose fields are the entity attributes that are physical parameters (e.g., ${\kappa},{\gamma},s$, resolution, width, etc) or attributes facilitating data management (e.g., computational time, file size, date and time of submission to the supercomputer job queue, etc).
For the complete list of attributes stored for each entity see Figure \ref{fig:ER}.

In a relational database, entities are related to each other by relationships.
For example, a `map' can {\it correspond to} a `macromodel', and a `macromodel' can {\it have} a `map' (the inverse relationship).
Additionally, each relationship can have a degree, e.g. a `map' can {\it correspond to} none or many `macromodel' entities i.e. a (0,N) relationship.
The GERLUMPH database entity types, relationships, degrees of relationships and primary keys are shown in the entity-relationship diagram of Figure \ref{fig:ER}.

The primary key can be one or more attributes unique to each entity.
In other words, each database table entry can be uniquely characterized by its primary key, or index.
We use this unique index of each entity to name the directory holding the actual data.

The contents (attributes) of each of the 6 GERLUMPH tables (entities), together with the file structure of the corresponding output directories, are described in the following.

\subsubsection{The `map' main table}
\label{sec:maptable}
The magnification map table contains all the meta-data relevant to a map.
Eight fields contain the map parameters viz. ${\kappa},{\gamma},s,N_{\rm avg}$, resolution, lens positions, number of lenses, and map width, which have been described in Sections \ref{sec:ray-shooting} and \ref{sec:data}.
There are four additional fields that contain the GPU computational time for generating the map, the GERLUMPH dataset the map belongs to (see Table \ref{tab:datasets}), the date and time of submission to the gSTAR job queue and the status of the submitted job (waiting in the queue, running, or complete).

The last field is the unique index of the map, which holds the name of the output directory containing the map related data.
Selecting maps from the database will seamlessly make use of the map index to retrieve the data.
All the data files ({\tt .dat} extension) are stored in text format and comprise about 0.02\% of the total data size.
The actual maps comprise the bulk of the GERLUMPH data, their data format and properties are discussed in more detail in \ref{app:compression}.
The data available in each directory consist of the following files:
\begin{enumerate}
\item {\tt map.bin}: ray counts per map pixel, the actual magnification map data, stored in 32-bit binary format. The size of this file is 381 Megabytes for our $10000^2$-pixel maps.
\item {\tt meta.dat}: input parameters used to generate each map and log information from gSTAR's queue. This compact file ($<$100 lines) contains information about the supercomputer nodes and GPUs used in the computation, as well as the number of microlenses used, the date and time of submission to the supercomputer, the status of the job, and the computational time, which are used to update the corresponding database table fields.
\item {\tt mapmeta.dat}: general physical properties of the map, stored separately for easy and quick access: ${\kappa},{\gamma}$, resolution, width, average number of rays per map pixel, ${\langle} N {\rangle}$, and average magnification, ${\langle} {\mu} {\rangle}$. Using the last two quantities one can convert from ray counts to magnification per pixel through equation (\ref{eq:mu}), where $N_{\rm avg}$ is given by the ratio ${\langle} {\mu} {\rangle} / {\langle} N {\rangle}$.
\item {\tt lenspos.dat}: input microlens positions in the lens plane.

\item {\tt icon.png}: an icon (200x200 sampled pixels) of the magnification map for preview purposes, stored as a Portable Netowrk Graphics\footnote{\tt http://www.w3.org/Graphics/PNG/} (PNG) image.
\item {\tt sample.png}: a larger image (1000x1000 sampled pixels) of the magnification map with more detail, stored in PNG format. This image is used by tools like {\tt Colorbar}, where a map preview is needed, but the map icon is not sufficiently large.
\item {\tt mpd.dat}: the full magnification probability distribution (MPD) of the map. For comparisons between maps, it is often more convenient to use the MPD, or magnification histogram; the frequency of each magnification value in a map is counted and then divided by the total number of pixels. An example map and the corresponding MPD are shown in Figures \ref{fig:getquery} and \ref{fig:colorbar}.
\item {\tt mpd.png}: a plot of the map MPD (680x510 pixels) for preview purposes, stored as a PNG image.
\end{enumerate}

\subsubsection{The `convolution' main table}
This table contains the convolution parameters viz. the unique index of the magnification map, the unique index of the accretion disc profile (see next section) and the value of $R_{\rm Ein}$ used.
Each convolution is assigned a unique index, which is used to name the output directory with the convolution results.
Finally, each entry contains the date and time of submission to the gSTAR job queue and the status of the submitted job.

Selecting maps (or disc profiles) from the database will seamlessly make use of the map (disc profile) and convolution index to retrieve all the related convolution data.
The data available in each directory consist of the following files:
\begin{enumerate}
\item {\tt convmeta.dat}: input parameters for each convolution and log information from gSTAR's queue kept as a reference. This is a small file ($<50$ lines) stored in text format.
\item {\tt mpd.dat}: the full magnification probability distribution of the convolved map, similar to the map MPD, stored in text format.
\item {\tt icon.png}: an icon (200x200 sampled pixels) of the convolved magnification map for preview purposes stored as a PNG image.
\item {\tt pixelFlux.bin}: typically a few thousand pixel locations and magnification values from the convolved map. This is a binary file, with pixel location on the map represented by two short integers (2 bytes each) and the magnification value by a float (4 bytes, see \ref{app:compression}).
\end{enumerate}

\subsubsection{The `disc' complementary table}
Keeping the disc profile information used for performing convolutions with GERLUMPH maps e.g., shape, size, etc, in a database table greatly facilitates generating and retrieving GERLUMPH results.
In this table, we store a short name for the profile (e.g., Gaussian${+}$hole), a long name, which better explains the profile properties (e.g., a two-dimensional normalized Gaussian distribution with a hole), the size of the profile in physical units, and additional information sufficient to reproduce the profile itself (e.g., the values for ${\sigma_{x}}$ and ${\sigma_{y}}$ of a two-dimensional Gaussian).
Finally, there is the unique profile index, which points to the output directory holding the profile data.

The disc brightness profile is calculated on a two-dimensional regular grid of 1000 to 3000 points per dimension, depending on the physical size of the profile.
The data available in each directory consist of the following files:
\begin{enumerate}
\item {\tt profileMeta.dat}: physical dimensions of the generated disc profile, and the resolution of the grid, $N_{x},N_{y}$, stored in text format.
\item {\tt profile.png}: an image of the disc profile (up to 3000x3000 resolution) for preview purposes, stored in PNG format.
\item {\tt profile.bin}: this is a binary file, with information stored as 32-bit float numbers.
The first $N_{x}$ and $N_{y}$ numbers hold the $x$ and $y$ values of the grid points in physical units.
The following $N_{x} {\times} N_{y}$ numbers represent the normalized brightness values in each grid location.
\end{enumerate}

At this stage, there are only a few sets of pre-computed disc profiles of some basic geometry in the `disc' table.
A tool could be envisaged here, allowing for users to upload their own accretion disc profiles as motivated by their work.
This would be considered in a future, fully integrated, online version of GERLUMPH and depend on the available hardware/software implementation (e.g. GPU-accelerated web-server, storing user data in the database, etc).
For the present, any additions/suggestions of accretion disc profiles should be addressed to the authors.

\subsubsection{The `macromodel' complementary table}
This table holds information on existing macromodels compiled from the literature originally by \cite{Bate2012}, and subsequent updates.
It provides the connection from specific macromodels and lens systems to the ${\kappa},{\gamma}$ parameter space.
Each entry consists of the name of the lens system (e.g., Q2237+0305), the name of the macromodel used (e.g., single isothermal sphere), the reference to the relevant publication, the link to the Astrophysics Data System\footnote{\tt http://adsabs.harvard.edu/index.html} (ADS) entry, the multiple image labels (e.g., A, B, etc), and the published ${\kappa},{\gamma}$ values for that image.

There is also a unique index for each entry, however, the corresponding output directory is empty.
This is because currently all the information required on macromodels can be stored entirely in the database.
However, the infrastructure is there in case more information on macromodels is required in the future.

A future possibility could be to integrate an online macromodeling tool \citep[e.g. Mowgli][]{Naudus2010} with the GERLUMPH magnification maps.
For the present, researchers that would like their macromodels specifically to appear in the GERLUMPH database are asked to contact the authors.

\subsubsection{The `job' management table}
Due to the very large number of map and convolution simulations that have to be performed (${\sim} 10^4$ and ${\sim} 10^6$ respectively), these are grouped in batches and submitted to the gSTAR job queue.
The `job' table holds information on batches of jobs, such as the version of the code used to carry out the simulations, the name of the gSTAR queue where the jobs were submitted (different queues have different GPU cards), the number of individual jobs in the batch, the unique job indices, a comment on the jobs (e.g., `GD2 maps part A'), the submission and completion date and time, and the status of the batch.
This information is useful in resolving issues which may occur with the supercomputer e.g. power/network failures, input/output errors, etc.

\subsubsection{The `download' management table}
\label{sec:download}
This table manages the interaction of users with the GERLUMPH data.
Currently, the option of downloading any of the GERLUMPH maps is enabled \citep[see also][]{Vernardos2014a}.
The `downloads' table contains a user identification code (no actual user data are stored, this is just a way to distinguish between different users), number of maps downloaded, the map indices, the compressed and uncompressed size of the data, the date and time of access and the status of the download.
The index of each entry points to a directory where the desired map data are temporarily stored in a compressed format ({\tt gzip} or {\tt bzip2}) and available for download.

Should the map download option be disabled in the future, the `download' table could still be used as a `user' table to manage a different kind of user interaction with the data.
This could include an inventory with the most commonly used maps, uploaded user-generated maps for comparisons, etc.

\section{eTools}
\label{sec:eTools}
Access to the GERLUMPH data requires simply a web browser and an internet connection to the server located at:
\begin{center}
{\tt http://gerlumph.swin.edu.au}
\end{center}
The web server uses the popular LAMP \citep[Linux Apache MySQL PHP, e.g. see][]{Lee2002} freely available, open-source software bundle.
The data managed by the web server are the magnification map data (${\sim}$25 Terabytes), the convolution data (${\sim}$5 Terabytes), and support files, such as preview icons, probability distributions, etc, (${\sim}$0.5 Terabytes) as described in Section \ref{sec:data}.
Users are expected to access the meta-data in the GERLUMPH database and the pre-computed support files most frequently.
Therefore, the provided set of online tools described below has been designed to make efficient and effective use of these data.

\begin{figure*}[!t]
\begin{center}
\includegraphics[width=\textwidth]{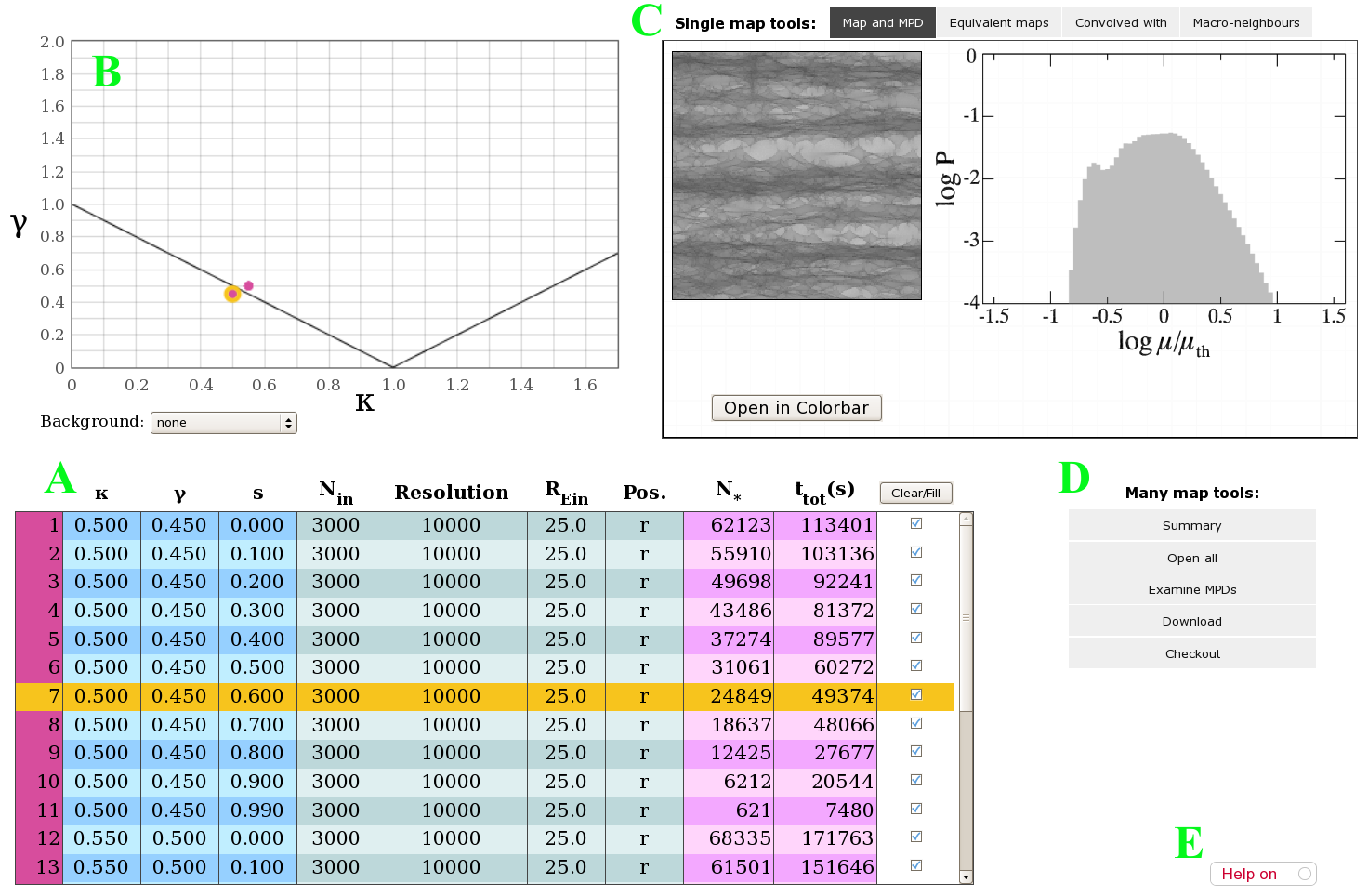}
\caption{
The main features of the {\tt getquery} tool, located at {\tt http://gerlumph.swin.edu.au/getquery/}:
\textbf{A} - Properties of the matching maps returned from a successful query to the `map' database table. The user can select any single map by clicking on the table rows, or a collection of maps by checking the boxes.
\textbf{B} - The ${\kappa},{\gamma}$ values of the selected maps. Properties of the parameter space can be displayed in the background. The user can click on the points to select single maps.
\textbf{C} - Tools for single maps: preview the map and its MPD, a list of equivalent maps, disc profiles the map has been convolved with, and macromodels that lie close to the map in the ${\kappa},{\gamma}$ parameter space.
\textbf{D} - Tools for multiple maps: a summary of the selected maps, previewing all the selected maps and MPDs, selecting the maps for download and proceeding to check out.
\textbf{E} - Activate help pop-up text for the current page, giving more details for each feature and tool.
}
\label{fig:getquery}
\end{center}
\end{figure*}

\begin{figure*}[t]
\begin{center}
\includegraphics[width=\textwidth]{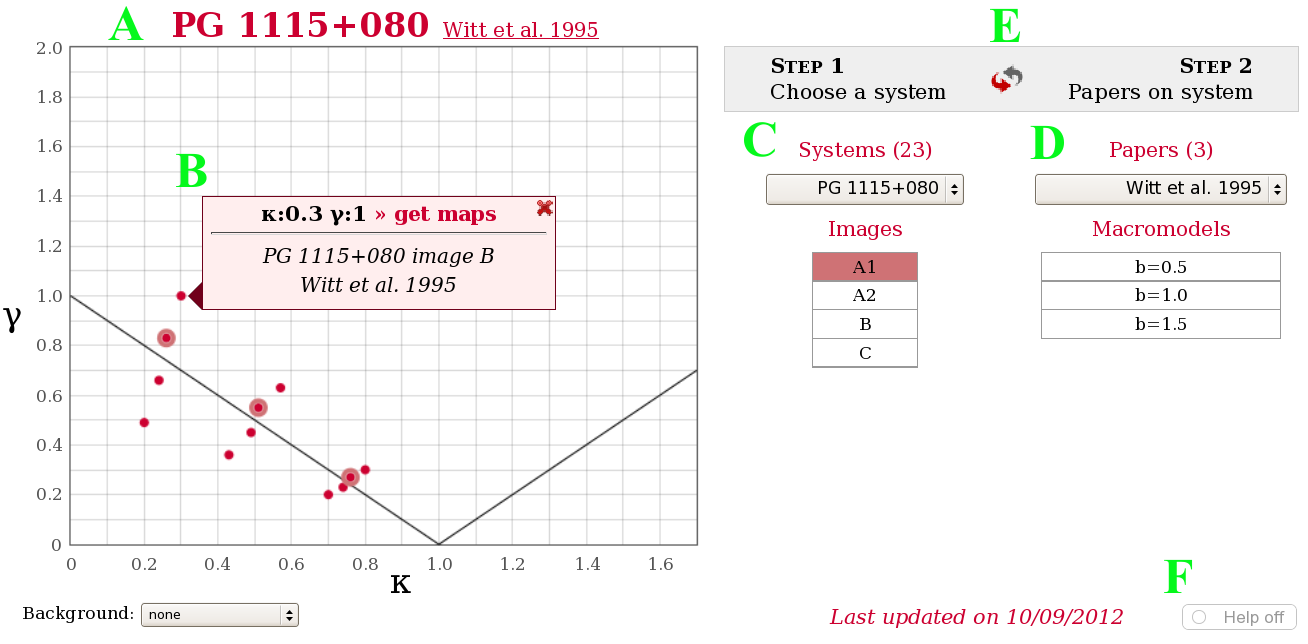}
\caption{
The main features of the {\tt macromodel} tool, located at {\tt http://gerlumph.swin.edu.au/macromodels/}:
\textbf{A} - The user can retrieve observational information by clicking on the selected system's name, or retrieve the publication abstract by clicking on the reference.
\textbf{B} - The ${\kappa},{\gamma}$ values of the selected system/macromodel. Properties of the parameter space can be displayed in the background. The user can click on the {\it get maps} link to retrieve the corresponding maps from the main `map' table.
\textbf{C} - Step 1: select a system from the database. A list of the multiple images of the system is displayed, which can be selected to highlight the corresponding values on the ${\kappa},{\gamma}$ plot.
\textbf{D} - Step 2: select a publication containing macromodels of the selected system (there may be only one paper available for some systems). A list of the macromodels is displayed, which can be selected to highlight the corresponding values on the ${\kappa},{\gamma}$ plot.
\textbf{E} - Switch the order of steps 1 and 2. For example, search for a given system and retrieve all the macromodels from the literature, or search for a given publication and retrieve all the macromodels it includes (for one or more systems).
\textbf{F} - Activate help pop-up text for the current page, giving more details for each feature and subsequent actions.
}
\label{fig:macro_tool}
\end{center}
\end{figure*}

To support online access to the GERLUMPH maps, we have developed a series of tools that can be used in an up-to-date web browser\footnote{recommended browsers are Firefox 17 and Chrome 32, or later versions.} with JavaScript support enabled.
The jQuery\footnote{\tt version 1.7: http://jquery.com/} JavaScript library has proven very useful in creating the tools described below.
For specific applications, we find that WebGL provides a performance boost, especially for online image processing (like map convolution).
WebGL is a JavaScript API for rendering interactive 3D and 2D graphics within any compatible web browser\footnote{to test whether your browser supports WebGL visit the following web page: {\tt http://get.webgl.org/}}, allowing for user-side GPU acceleration.
In the following, we describe ways of accessing and analyzing the GERLUMPH data through a set of online tools.

\subsection{Query the map table: the {\tt getquery} tool}
The {\tt getquery} tool is the basic tool providing general access to the GERLUMPH datasets (Section \ref{sec:data} and Table \ref{tab:datasets}) via the database tables described in Section \ref{sec:databases}.

\subsubsection{Description}
A range of values for ${\kappa},{\gamma}$ and $s$ can be specified, and the matched results from any of the GERLUMPH datasets will be returned.
The properties of the resulting maps are listed in a table, and their ${\kappa},{\gamma}$ values are plotted in the parameter space.
Parameter space results from \cite{Vernardos2013} and \cite{Vernardos2014a}, along with other properties of the ${\kappa},{\gamma}$ parameter space (e.g. the number of microlenses, magnification contours, etc), can be displayed in the background of the plot.
Online tools are provided for further analysis of single, or groups of maps.
A screenshot highlighting the main features of the {\tt getquery} tool is presented in Figure \ref{fig:getquery}.

Single maps can be selected by clicking on a table row, or a point on the parameter space plot, while multiple maps can be selected by checking the boxes on the table. 
The single map tools are then automatically updated by loading a map and MPD preview, a set of equivalent maps due to the mass-sheet degeneracy, a list of all the accretion disc profiles the map has been convolved with, and a list of neighbouring macromodels of actual systems on the ${\kappa},{\gamma}$ parameter space.
Further actions and links to specialized tools are provided for more detailed analysis e.g. the map preview is linked to the {\tt colorbar} tool (see Section \ref{sec:colorbar}) for caustic structure inspection, the list of equivalent maps and neighbouring macromodels can be plotted on the ${\kappa},{\gamma}$ parameter space, etc.
The available multiple-map tools consist of a summary of the selected maps (number of maps, size of data, etc), previewing all the maps and MPDs, selecting the maps for download and proceeding to check out.
Each of these actions is accompanied by detailed instructions, either by enabling the help instructions (location E on Figure \ref{fig:getquery}), or in the description found at the top of each tool's own web page.

\subsubsection{Implementation}
All the tools are based on a template: a description of each tool's functions and features appears at the top of the web page, followed by the main part of the tool, where all the user interactions take place.
The tool web page is loaded by a template PHP script, and then the content is initialized by a JavaScript function.
In the case of the {\tt getquery} tool, this content can be the default ${\kappa},{\gamma},s$ values, or other initial values, in the case where the user is redirected here from another tool (e.g., the {\tt macromodel} tool, see Section \ref{sec:macrotool}, or the {\tt getquery} tool itself, see below).
Further calls to other PHP scripts may result, which perform specialized tasks as described below.

Upon changing the values for the ${\kappa},{\gamma},s$ and selecting the GERLUMPH dataset in the relevant HTML form fields, a request will be sent to the server.
The request is handled using AJAX\footnote{Asynchronous JavaScript and XML (AJAX) is not another programming language, rather a new way to use existing standards: {\tt http://www.w3schools.com/ajax/}} techniques, which allow updates to parts of a web page without reloading the whole page.
The main part of the tool (shown in Figure \ref{fig:getquery}) is updated with a list of matching GERLUMPH maps, returned by a PHP script that performs a query on the ${\kappa},{\gamma},s$ and dataset fields of the `map' database table.
Once the response from the server is received, JavaScript callback functions create a HTML table listing the results, and plot the ${\kappa},{\gamma}$ values in the parameter space.
The parameter space plot is drawn using Flot\footnote{\tt http://www.flotcharts.org/}, a JavaScript plotting library.

Selecting a single map from the table (click on a row), or from the parameter space plot (click on a point), will launch a series of AJAX calls, which will update the content of the single tool panels (location C on Figure \ref{fig:getquery}) without reloading the web page.
The following results are returned by PHP scripts: the map and MPD preview icons, a list of equivalent maps due to the mass--sheet degeneracy \citep[see][]{Vernardos2014a}, a list of neighbouring macromodels from the literature (if any), and a list of accretion disc profiles the map has been convolved with.
The only information required by these scripts is the map unique index, which is a hidden field in the HTML table, already available from the previous communication with the database.
The returned information is either metadata from the database tables (e.g. equivalent maps are returned from the `map' table, neighbouring macromodels from the `macromodel' table, accretion disc profiles from the `disc' table, etc), or actual data from the output directories (e.g. the {\tt icon.png} and {\tt mpd.png} preview icons).

Further tasks can be performed by each single map tool: the image of the map can be inspected in more detail than the preview icon using the {\tt Colorbar} tool (Section \ref{sec:colorbar}), the equivalent maps can be overlaid on the ${\kappa},{\gamma}$ parameter space, the collection of equivalent maps can be loaded separately in the {\tt getquery} tool, and existing neighbouring macromodels can be inspected using the {\tt macromodel} tool (Section \ref{sec:macrotool}).
These tasks are either performed on the current web page by JavaScript functions using data already loaded from the server, or by passing on the necessary data (usually just the unique indices of maps, macromodels, etc) to other tools, opening a new web page.

The tools for multiple maps make use of the unique indices of the checked maps in the HTML table.
A summary of the selected maps i.e. number of maps, total uncompressed data size and set of parameters, can be displayed by a JavaScript function using existing loaded information from the server.
The preview icons of all the maps and their corresponding MPDs can be opened in a separate web page, where the map indices are passed by simple form submission using the HTML GET or POST methods.

When the {\it Download} option is chosen, the currently selected map indices are stored in a PHP session array variable.
The values in this array are available to any page within the GERLUMPH website.
In this way, when the user proceeds to {\it Checkout}, the contents of this PHP session variable are displayed, which include all the maps selected for download while browsing with the GERLUMPH tools.
Moreover, each user session is assigned a unique identification code by the server, which is the one used to distinguish between users in the `download' database table (see Section \ref{sec:download}).
This is the only application that uses PHP session variables on the server.

\subsection{Specific systems: the {\tt macromodel} tool}
\label{sec:macrotool}
GERLUMPH magnification maps for specific multiply imaged systems can be accessed by using the {\tt macromodel} tool.

\subsubsection{Description}
The existing macromodels from the literature, as compiled by \cite{Bate2012}, have been inserted in the `macromodel' database table, according to the schema shown in Figure \ref{fig:ER}.
The user can proceed in two interchangeable steps: select a specific system and retrieve all its macromodels and relevant publications, or select a specific publication, which can contain more than one system or macromodel.
Matching database entries are returned and plotted on the ${\kappa},{\gamma}$ parameter space.
The plotted ${\kappa},{\gamma}$ values can then be highlighted according to which multiple image or macromodel (as presented in the related publication) they belong to.
Finally, links to the publication abstracts are provided and an image of the system is displayed (usually from CASTLES), together with lens and source redshifts, and other observed properties from the literature.
A screenshot outlining the features of the {\tt macromodel} tool described above is shown in Figure \ref{fig:macro_tool}.

The connection to the GERLUMPH magnification map database is provided by clicking on any of the ${\kappa},{\gamma}$ points on the plot.
A short description appears (location B in Figure \ref{fig:macro_tool}) with information about the system, image, macromodel, and publication the selected ${\kappa},{\gamma}$ combination corresponds to.
A link to the {\tt getquery} tool page is provided, which will automatically perform a query to the `map' database and return all the available maps for the specific macromodel in question.
After retrieving the maps, one can proceed further as described in the {\tt getquery} tool section.

\subsubsection{Implementation}
A description of the {\tt macromodel} tool and its features appears at the top of the web page, followed by the main part of the tool shown in Figure \ref{fig:macro_tool}, where all the user interactions take place.
The web page is generated by a PHP script, and then the content is initialized by a JavaScript function to display all the available systems in the first step.
If a user is redirected here from the {\tt getquery} tool, the {\tt macromodel} tool will be initialized with the selected system.
Modifying the available options launches a series of AJAX calls to PHP scripts, which retrieve various pieces of information from the `macromodel' database table.
Subsequently, the content of the main part of the tool will be updated by JavaScript callback functions, without reloading the web page.

This tool works in two consecutive steps: start with a given system and retrieve all the related papers, or start with a paper and retrieve all the systems it includes.
For a selected system in step one, the list of papers on that system will be updated in step 2, the labels of the multiple images of the system will be loaded and displayed, and all the macromodel ${\kappa},{\gamma}$ values for that system will be plotted in the parameter space.
If there are more than one papers on the selected system, another series of AJAX calls will be performed when one of the papers is selected.
Alternatively, the available macromodels and the link to the ADS abstract (appearing above the parameter space plot) will be updated straightaway.

Observational properties on a selected system are also loaded, but remain hidden in the background until the user selects to inspect them by clicking on the system's name appearing on the top of the parameter space plot.
The associated PHP script retrieves an image of the system by searching the CASTLES web page, using the default XML parser to read its contents, and gets additional observational information, such as the lens and source redshifts, the $R_{\rm Ein}$, etc, from Table 1 of \cite{Mosquera2011b}, stored locally as a text file.
This is a basic implementation of getting observational information from different available sources, however, a connection to more extensive databases, like the MasterLens\footnote{\tt http://masterlens.astro.utah.edu/} project, could be envisioned here.

\subsection{Mean MPD}
Apart from the general {\tt getquery} and {\tt macromodel} tools, that can be used with any of the GERLUMPH maps, there are online tools that are specially designed for particular GERLUMPH datasets.
One such tool is the {\tt meanMPD} tool, which is designed specifically for GD0 maps.

\subsubsection{Description}
\begin{figure}[t]
\begin{center}
\includegraphics[width=0.47\textwidth]{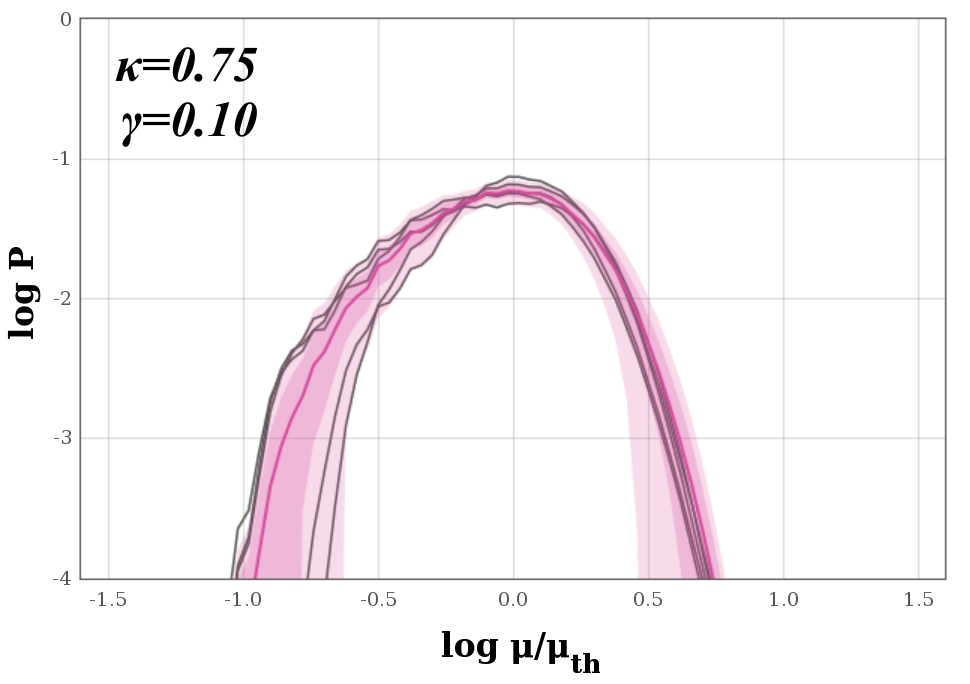}
\caption{An example of the {\tt meanMPD} tool, which is using the GD0 maps. The mean MPD (magenta line) and its ${\pm}2{\sigma}$ deviations are plotted. Five individual MPDs (grey lines) are also shown. By comparing them to the mean, one can see that two MPDs are decreasing faster than the rest, for low magnification values. The {\tt meanMPD} tool is accessible at {\tt http://gerlumph.swin.edu.au/tools/meanMPD/}.}
\label{fig:mean}
\end{center}
\end{figure}

This tool shows variations of the MPD between 15 different realizations of the same map, having the same parameters ${\kappa},{\gamma}$, etc, but different microlens positions.
The derived mean MPD and standard deviation, which have been presented in Figure 4 of \cite{Vernardos2013}, can also be displayed.
Using this tool gives more control over these results, as one can choose which individual MPDs to display, as well as the number of bins used for the distributions.
In this way, individual MPDs that are quite different from the mean can be identified and understood.
An example is shown in Figure \ref{fig:mean}, for a trial combination of ${\kappa},{\gamma} = (0.75,0.1)$.

\subsubsection{Implementation}
The ${\kappa},{\gamma}$ values are selected in a similar way in all the tools for specific datasets: a parameter space plot is created with all the available values from the targeted GERLUMPH dataset, the user can browse through this plot and select different ${\kappa},{\gamma}$ combinations.

An AJAX call executes a PHP script on the server returning the desired data and/or metadata.
The {\tt meanMPD} tool is using data from the {\tt mpd.dat} file (see Section \ref{sec:databases}) that holds the magnification probability distribution.
However, these data are binned into a histogram before they are returned to the user.
By default, the number of bins is set to 100, with a maximum limit of 400 that can be selected by the user.
The returned histogram data are then stored in JavaScript variables and can be further manipulated.

After the individual and mean MPD histogram data are returned from the server, a JavaScript callback function plots the probability distributions.
There are 15 individual MPDs for each ${\kappa},{\gamma}$ combination from GD0, the derived mean MPD, and the ${\pm}1{\sigma}$ and ${\pm}2{\sigma}$ regions, which can be toggled on the plot by selecting checkboxes that correspond to each of the histograms.

\subsection{MPDs}
This tool examines the effect of smooth matter on the MPDs and is designed specifically for the GD1 dataset.

\subsubsection{Description}
\begin{figure*}[t]
\begin{center}
\includegraphics[width=\textwidth]{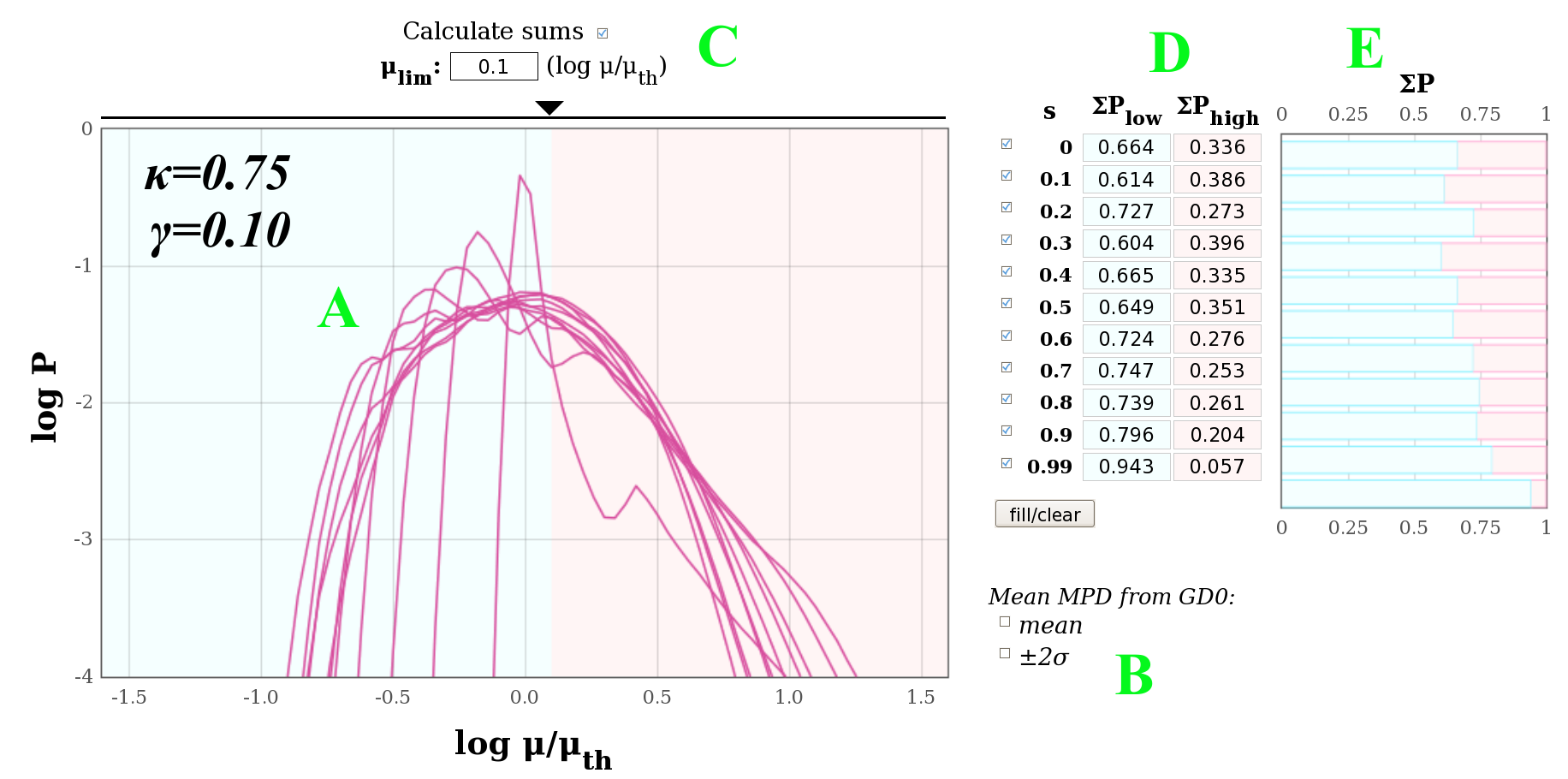}
\caption{
The main features of the {\tt MPDs} tool, which is using the GD1 data and is accessible at {\tt http://gerlumph.swin.edu.au/macromodels/}:
\textbf{A} - The individual MPDs, for different values of $s$.
\textbf{B} - The mean MPD and standard deviation can be displayed, whenever available (from GD0, for $s=0$).
\textbf{C} - The user can type-in, or use the slider, to select a value for ${\mu_{\rm lim}}$. The probability sums are then calculated and plotted as a function of $s$. For this example, ${\mu_{\rm lim}}$ has been set to 1.25 ${\times} {\mu_{\rm th}}$ (log ${\mu}$/${\mu_{\rm th}} = 0.1$).
\textbf{D} - The calculated probability sums are displayed here. Displaying the MPDs can be controlled by the checkboxes.
\textbf{E} - Plot of the probability sums as a function of $s$. The sum for ${\mu} < {\mu_{\rm lim}}$ is shown in light blue and the sum for ${\mu} > {\mu_{\rm lim}}$ is shown in light red. The total probability sum for all values of ${\mu}$ has to be equal to unity.
}
\label{fig:mpds}
\end{center}
\end{figure*}

A plot showing variations of 11 MPDs for a given ${\kappa},{\gamma}$ pair, but different $s$ values, is created.
The mean MPD and standard deviation  can also be displayed whenever available (from GD0, only for $s=0$).
Probability sums for ${\mu} < {\mu_{\rm lim}}$ and ${\mu} > {\mu_{\rm lim}}$, where ${\mu_{\rm lim}}$ is a magnification value set by the user, can be calculated and plotted as a function of $s$.
In this way, the effect of including smooth matter on microlensing induced magnification fluctuations can be examined.
In Figure 6 of \cite{Vernardos2014a}, the values of the probability sums for ${\mu} < 0.3 {\mu_{\rm th}}$ and ${\mu} > 3 {\mu_{\rm th}}$, where ${\mu_{\rm th}} = {\frac{1}{(1-\kappa)^2-\gamma^2}}$ is the macro-magnification, are investigated as a function of $s$.
Using the {\tt MPDs} tool, probability sums can be calculated and similar plots can be produced, up to (or beyond) any value of magnification.
A screenshot outlining the features of this tool is shown in Figure \ref{fig:mpds} for a trial combination of ${\kappa},{\gamma} = (0.75,0.1)$, for which the mean MPD is shown in Figure \ref{fig:mean}.

\subsubsection{Implementation}
The {\tt MPDs} tool has been implemented almost identically to the {\tt meanMPD} tool.
The nature of the data returned from the server is the same: a collection of MPDs binned into histograms.
However, there is the additional feature of calculating probability sums once the data are returned to the user and stored in JavaScript variables.
The ${\mu_{\rm lim}}$ for the sums is set using a text input field, or a slider, generated using the jQuery user interface\footnote{http://jqueryui.com/} (location C in Figure \ref{fig:mpds}).
The background of the plot is actually a HTML5 canvas element, which can be drawn using JavaScript functions creating the colored background in location A of Figure \ref{fig:mpds}.
Once the sums for low and high probabilities are calculated as a function of $s$, a barplot is created using Flot (location E in Figure \ref{fig:mpds}).

\subsection{Probability surface}
The {\tt P-surface} tool shows variations of the probability surface with respect to magnification and smooth matter fraction.

\subsubsection{Description}
\begin{figure}[t]
\begin{center}
\includegraphics[width=0.47\textwidth]{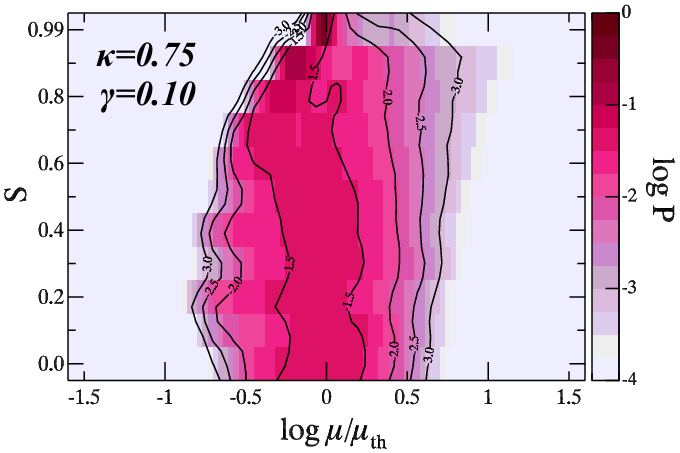}
\caption{The probability surface for ${\kappa},{\gamma} = (0.75,0.1)$, as a function of $s$ and ${\mu}$. Contours are drawn at log P = -1.5,-2,-2.5,-3.5, and can be controlled by the user. There is a gradual narrowing of the MPD as $s$ is increased, and a secondary peak can be observed for $s=0.7-0.8$.}
\label{fig:psurface}
\end{center}
\end{figure}

The same MPD information as in {\tt MPDs} is used, but visualized in a different way.
The changes in the shape of the MPDs with respect to $s$ are more clearly seen on the probability surface, and contours are used to further highlight its shape.
In Figure 4 of \cite{Vernardos2014a}, similar probability surfaces are displayed, however, with this tool the user has control over the contour levels and the number of bins used for the MPDs.
An example probability surface for a trial combination of ${\kappa},{\gamma} = (0.75,0.1)$ is shown in Figure \ref{fig:psurface}.
The individual MPDs for this ${\kappa},{\gamma}$ combination are shown in Figure \ref{fig:mpds}, and the mean MPD for $s=0$ is shown in Figure \ref{fig:mean}.

\subsubsection{Implementation}
Exactly the same binned MPD data as in the {\tt MPDs} tool are used for creating the probability surface plot, however, they are not returned to the user's browser and all the processing happens on the server side.
The surface is plotted with the user specified contour levels using the Python matplotlib\footnote{http://matplotlib.org/} library.
A PNG image of the plot is produced and returned to the user's browser.


Processing of the probability surface data on the user's side could be performed with a suitable JavaScript algorithm that calculates contours based on the data.
Extensions of such contour plots to larger sets of data and/or a real time contour drawing could be achieved by porting the contour algorithm to WebGL.
Developping such an algorithm is beyond the scope of this work.

\subsection{The {\tt colorbar} tool}
\label{sec:colorbar}
Characteristics of the caustic network on a map and features of its corresponding MPD can be highlighted using the {\tt colorbar} tool.
In doing so, properties which may not be directly noticeable in a map with fixed colors can be emphasized and studied.

\subsubsection{Description}
\begin{figure*}[t]
\begin{center}
\includegraphics[width=\textwidth]{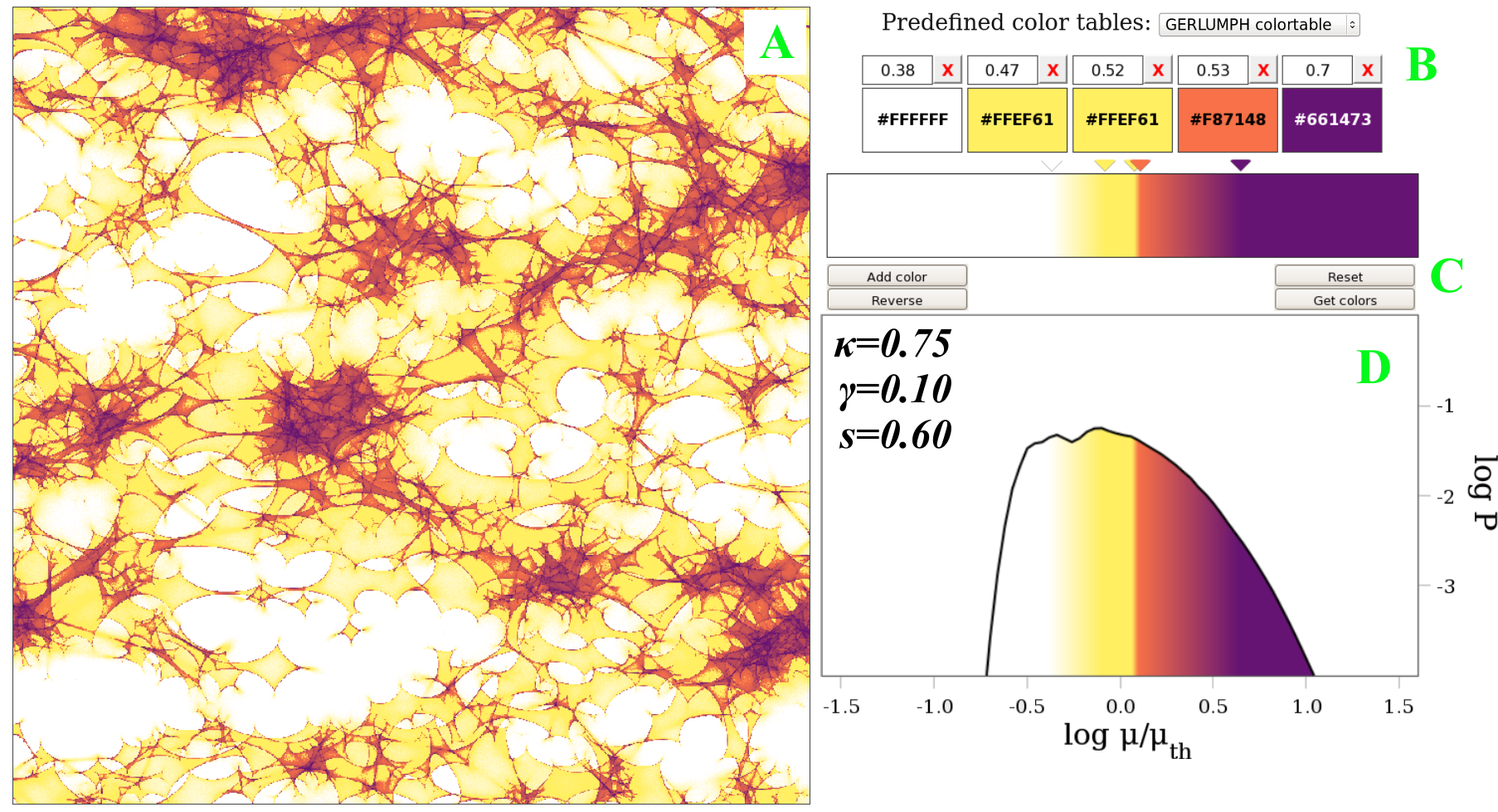}
\caption{
A user created color bar and the corresponding regions on the magnification map and MPD. Yellow and white colored regions on the map correspond to ${\mu} < 1.25 {\times} {\mu_{\rm th}}$ magnification values (the yellow color has been used twice to create a sharp transition between yellow and orange in the colorbar).
The {\tt colorbar} tool is located at {\tt http://gerlumph.swin.edu.au/tools/colorbar/} and its main features are:
\textbf{A} - A 1000x1000 pixel map sample, showing the networks of caustics colored according to the color bar.
\textbf{B} - A user created color bar, with the options of selecting the colors and changing their position on the bar.
\textbf{C} - Options for manipulating the colorbar and downloading its RGB values in text format.
\textbf{D} - The corresponding MPD, colored according to the color bar.
}
\label{fig:colorbar}
\end{center}
\end{figure*}

An example of the {\tt colorbar} tool can be seen in Figure \ref{fig:colorbar}, for $\kappa,\gamma,s = (0.75,0.1,0.6)$.
Magnification values for which log ${\mu}$/${\mu_{\rm th}} > 0.1$  ($1.25 {\times} {\mu_{\rm th}}$) are shown in orange and purple colors.
The probability to get a magnification value coming from these regions of the map is 0.276 (see Figure \ref{fig:mpds}, where the probability sum is calculated for these values of ${\kappa},{\gamma},s$ and for log ${\mu}$/${\mu_{\rm th}} > 0.1$).

It is very hard to display and manipulate a full high resolution map in our online approach: displaying a $10000^2$ pixel map in full resolution would require 6${\times}$9 HD resolution monitors (1920${\times}$1200 pixel resolution) while manipulating the color of its pixels in real time may require more than one GPU.
{\tt Colorbar} is designed to allow for easy map visualization in a web browser.
Therefore, we are using a $1000^2$-pixel sample image of each map for this purpose (the {\tt sample.png} file discussed in Section \ref{sec:databases}).
This image can be easily scaled on a standard desktop or notebook display with vertical resolution of 800 pixels.

It is underlined here that the purpose of the $1000^2$--pixel map sample and the {\tt colorbar} tool is to provide an easy and fast way of exploring the two-dimensional configuration of caustics and their connection to the MPD; it is more of a demonstration tool.
We find that sampling the full map is sufficient for this purpose; selecting a small number of pixels to represent the whole map, without using information from their neighbouring pixels, as opposed to the slower method of rescaling the map with some pixel averaging.
Any scientific goal, such as map convolution or light curve extraction, should make use of the full high resolution GERLUMPH maps.

\subsubsection{Implementation}
The colorbar itself (location B in Figure \ref{fig:colorbar}) is a HTML5 canvas element, drawn using standard JavaScript methods to create and apply linear gradients.
A list of predefined colorbars is available as an option\footnote{color symbols from {\tt http://colorbrewer2.org/}}.
The colors used in the gradient appear above the colorbar and can be added, removed or modified by the user.
The position of the colors in the gradient can be typed in, or selected with the sliders (created using the jQuery user interface).
Color values can be typed in using a hexadecimal (hex) notation for the combination of Red, Green, and Blue color values (RGB), or selected via the JSColor\footnote{\tt http://jscolor.com/} color picker.

The MPD data of the selected map are loaded upon initialization, and binned in a 100-bin histogram.
The JavaScript functions used to plot the probability distributions (location D in Figure \ref{fig:colorbar}) are the same as in the {\tt meanMPD} and {\tt MPDs} tools.
However, the area under the plotted data is set to transparent, allowing for the background color gradient to be visible, which is simply another canvas element drawn exactly as the colorbar, thus creating the effect of coloring the MPD.

The map pixel data are loaded upon initialization from the {\tt sample.png} file.
In fact, the colored map (location A in Figure \ref{fig:colorbar}) is simply another canvas element drawn by WebGL.
After the colorbar has been initialized and the data retrieved from the server, the pixel values from the map and the colorbar canvas are loaded into WebGL textures.
The map sample is a black and white image loaded as a two--dimensional texture, where each pixel has the same RGB color triplet, scaled within the allowed range of colors (0-255).
The pixels from the colorbar canvas are read in a one--dimensional texture of 256 pixels, each with the RGB triplet of the corresponding color on the colorbar.

Most of the WebGL related set-up takes place using JavaScript, except the part of the code which will run on the GPU viz. the vertex and fragment shaders, written in GLSL\footnote{OpenGL Shading Language, for more information visit the web page: {\tt http://www.khronos.org/opengles/} }.
The WebGL--drawn scene is very simple -- a single square object filling the entire field of view -- meaning that the vertex shader used to create it is very basic.
The fragment shader is using the loaded textures to color the map pixels: the grayscale value of each map pixel (0-255) is used as an index for the colorbar pixel texture, using the retrieved RGB triplet from the colorbar to assign its color to the map pixel.

The process of assigning colors to map pixels could be performed using JavaScript only, which runs on the CPU, but there would be severe performance penalties.
Our WebGL implementation is using the GPU, allowing large number of pixels to be processed in parallel, and creating the real time effect of changing the map colors.

\section{Discussion}
\label{sec:discussion}
GERLUMPH is a cosmological microlensing theoretical parameter survey that takes up the challenge of parameter space exploration, in preparation for the new discoveries of microlensed quasars by synoptic all--sky surveys of the next decade.
The goal is to understand the structure of the central quasar regions, mainly the supermassive black hole and the surrounding accretion disc, making connections to quasar and galaxy evolution in the Universe.

A number of applications to using the GERLUMPH results have been presented in the form of online eResearch tools, described in Section \ref{sec:eTools}.
The total size of the GERLUMPH data is ${\sim}$30 Terabytes (see Section \ref{sec:data}), which is managed by a database and a web interface (see Section \ref{sec:databases}).
We comment below on our approach to turning such a moderately sized dataset into a resource open to the community, the flexibility of the components we decided to use, and potential future extensions.

The first step was to use freely available open source software.
The nature of our application turned out to be within the capabilities of such software, without the need for optimized commercial solutions; the size of the database (70,000 maps and ${\sim}$10$^6$ convolutions) and the basic nature of the queries performed are perfectly manageable by the open version of MySQL.
Moreover, using popular solutions within the community of web developpers (e.g., the LAMP scheme), means that there is a plethora of available libraries and extensions, as well as extensive documentation.
Finally, experimenting with new technologies, such as WebGL, provided us with insight on developing powerful online tools for astronomy, especially regarding image processing.

The three main components of the GERLUMPH online resource are the data, database, and web servers (see Figure \ref{fig:flow}).
Our current implementation is using separate physical machines for each of those components, located at and supported by Swinburne University of Technology.
However, it is straightforward to move the components to other academic, or commercial, service providers, with minimal changes to the scripts.
Such action is not presently required, but it may be considered in the future, after taking the community needs and usage into account.

The initial set of online tools described in this paper has been designed to demonstrate the power and flexibility of our data and database implementation.
One can select and inspect properties of any ${\kappa},{\gamma}$ range of interest within the GERLUMPH datasets.
As an example, for ${\kappa}=0.75$ and ${\gamma} = 0.1$ one can:
\begin{itemize}
\item examine the dependence of the MPDs on the random positions of microlenses (Figure \ref{fig:mean});
\item examine the dependence on $s$ and calculated probability sums in Figure \ref{fig:mpds}, the shape of the probability surface (Figure \ref{fig:psurface}); and
\item inspect the caustic networks on the actual maps and their connection to the MPD (Figure \ref{fig:colorbar}).
\end{itemize}
It is straightforward to directly compare the output of the online tools to existing parameter space results \citep[e.g.][]{Vernardos2013,Vernardos2014a}.
Further connections to macromodels, observational data and convolution results have been demonstrated.

Future uses of the GERLUMPH infrastructure can be envisioned:
\begin{description}
\item {\bf Observer's tools}.
With the currently available GERLUMPH convolution data, theoretical flux ratios could be extracted online and used in comparisons with existing \citep[e.g.][]{Bate2008,Floyd2009,Jimenez2014} or future observations.
A similar online tool could be envisioned for generating light--curves.

\item {\bf Server side processing on request or WebGL?}
Instead of pre--computing all the possible map--profile combinations on gSTAR and storing them in the database for users to access, we could allow for this process to take place on the server on user request.
This would likely require a dedicated high--end server, possibly using GPU acceleration at various stages of data generation, and a scheduler to manage the user requests.
However, our GPU--accelerated convolution code is reasonably fast (${\sim}$3 s for a convolution of a 10000$^2$--pixel map with a disc profile), with further possibilities for additional speed--up (e.g. light--curve extraction on the GPU), making a cost--effective solution realistic.

Should such higher level user interactions with the server and data be enabled, it would also be possible to allow for users to upload and store their own generated magnification maps.
Using a common predefined data format, all the platform of GERLUMPH tools could be used on the uploaded user maps.
The likely amount of maps uploaded will be manageable by the database, and only sufficient storage disk space should be provided.

The possibility of performing all these tasks on the user's side by using high performance WebGL tools should be investigated.
The outcome would be to distribute the demanding computations from the server to the users' GPUs, without the need for a powerful server machine, but with the trade--off being much larger size of data being transfered from and to the server and database.

\item {\bf Outreach and education}.
The use of web technologies with astronomical data by the public has proven to be successful in a number of cases \citep[e.g. Galaxy zoo,][etc]{Raddick2010}.
A set of tools specially designed for outreach and education purposes could be imagined, with the first candidate being the implementation of the inverse ray--shooting technique itself (\ref{app:release}) in WebGL.
An advantage of a web--based service is that an alternative front--end can be applied depending on the specific needs of the users: from novice to master user.
\end{description}

\section{Conclusions}
\cite{Thompson2010} and \cite{Bate2010} demonstrated the benefits of combining the inherent parallelism of the inverse ray-shooting technique with a GPU graphics card, resulting in the {\tt GPU-D} code.
On the basis of these results,  \cite{Bate2012} suggested a strategy for a cosmological microlensing theoretical parameter survey that would require a GPU supercomputer, and GERLUMPH is the realisation of this vision.

Since the inauguration of the gSTAR supercomputer at Swinburne University of Technology in 2012, we have made rapid progress on GERLUMPH: more than 70,000 high-resolution magnification maps are already available for microlensing studies.   The four GERLUMPH datasets (GD0-GD3) comprise a public, freely available data resource.

Investigating properties of the microlensing parameter space is a crucial step in preparation for the upcoming synoptic all--sky survey era \citep[see][for results and discussion]{Vernardos2013,Vernardos2014a}.
With the imminent inflow of data for thousands of multiply imaged systems, the GERLUMPH online resource provides the tools for supporting future studies on statistically interesting samples of microlensed quasars.
This includes both quick look--up of the microlensing properties for each new discovery, or more extensive analyses across all of microlensing parameter space.

Our online, browser-based eTools encourage analysis to occur via the remote processing paradigm.
It is straightforward to add additional eTools driven by the needs of the quasar microlensing community.
Through integration with the existing CASTLES gravitational lens database and links to publications via the ADS abstract service, the GERLUMPH eTools provide a natural connection between observational results and theoretical simulations.
Together with next--generation observational databases (e.g. MasterLens) and other modelling applications \citep[e.g. Mowgli][]{Naudus2010}, the GERLUMPH resource provides a key pillar of the future microlensing eResearch cloud.

\section*{Acknowledgments}
This research was undertaken with the assistance of resources provided at gSTAR through the ASTAC scheme supported by the Australian Government.
gSTAR is funded by Swinburne and the Australian Government's Education Investment Fund.
The authors would like to thank Alex Thompson, Nick Bate and Darren Croton.
C.J.F. is grateful to NVIDIA for providing the K20 and K40 cards through the NVIDA CUDA Research Center program.
This research has made use of NASA's Astrophysics Data System Bibliographic Services.

\bibliographystyle{apalike}
\bibliography{etools}

\appendix
\section{{\tt GPU-D}: code release}
\label{app:release}
Graphics processing units (GPUs) have emerged as credible, low-cost, computational co-processors, capable of providing speed-ups from ${\mathcal{O}}(10)$ to ${\mathcal{O}}(100)$ compared to low-core-count CPU codes.
The massively parallel architecture of GPUs means that they provide the best performance for algorithms with high levels of data parallelism and high arithmetic intensity (i.e. a high ratio of floating point computations compared to memory look-ups).

\begin{figure}
\begin{center}
\includegraphics[width=7.6cm,angle=0]{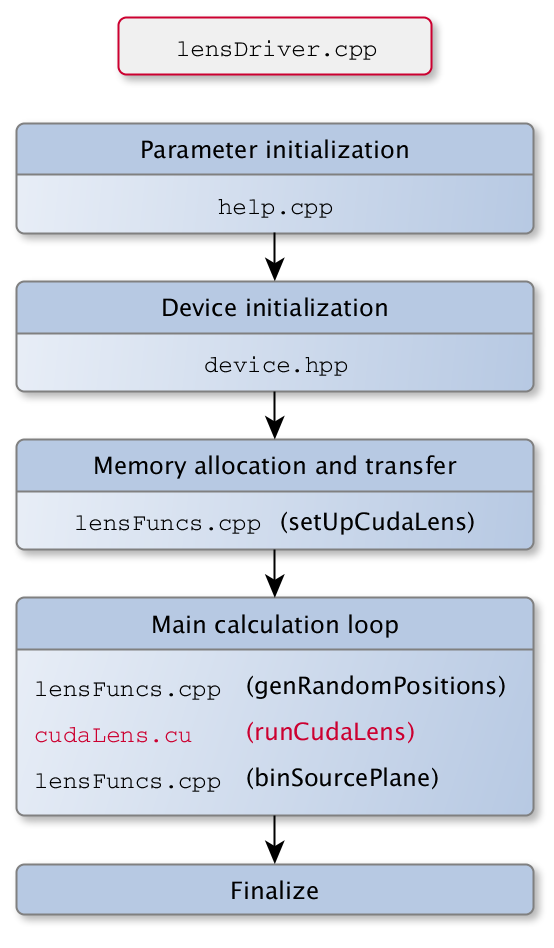}
\caption{A schematic control flow of the {\tt GPU-D} code. The main program {\tt lensDriver.cpp} initializes the parameters, initializes the GPU device, allocates and transfers memory between the CPU and the GPU, and executes the main calculation. The CUDA kernel that directly solves equation (\ref{eq:lenseq}) in parallel is in the {\bf runCudaLens} function, in the {\tt cudaLens.cu} file.}
\label{fig:gpud_flow}
\end{center}
\end{figure}

Although the parallelisation of GPU--based brute force ray--shooting is discussed at length in \cite{Thompson2010}, the code itself was not released.
We now remedy this by releasing a version of {\tt GPU-D} that has the basic functionality for producing magnification maps.
The important part is the CUDA kernel, not what the code is wrapped inside (see Figure \ref{fig:gpud_flow}).
The {\tt GPU-D} code can be downloaded from:
\begin{center}
{\tt http://gerlumph.swin.edu.au/GPUD/}
\end{center}
and is available in the Astrophysical Source Code Library \citep[ASCL;][]{Thompson2014}

Through this code release, we encourage additions and enhancements to the core functionality, along with providing opportunities for others to benchmark this code against their own preferred alternatives.
Due to slight differences in system architecture, and the idiosyncrasies of GPUs (e.g. computations are not necessarily associative, depending on the order in which the elements of a summation are scheduled), the exact output generated by other users may differ subtly from results in the GERLUMPH database.

\section{The data format and compression}
\label{app:compression}
The bulk of the GERLUMPH data consists of magnification maps, amounting to a total of ${\sim}$25 Terabytes in size.
The map data format and compression scheme will effect users attempting to download maps from the web server (see Section \ref{sec:eTools}) to use with their own analysis tools.
Moreover, the total size affects the way data is stored; for the moment, map data reside on gSTAR's hard drives, which have a total capacity of ${\sim}$1 Petabyte, more than enough for GERLUMPH's needs.

Nevertheless, GERLUMPH was designed to be a flexible resource, and although currently there is no plan to move off gSTAR, this may be considered in the future (e.g., moving to a dedicated data and web server, or a cloud-based solution).
In this case, the total data size will play a role and compression options will have to be considered.
In this section, we discuss the chosen data format and techniques of compression for the magnification map files.

From a data point of view, a map is just a series of integer values (ray-counts per pixel).
The maximum pixel value in a map will set the number of bits used to store its pixel ray-counts, which in turn will set the final file size.

The first step, is to find the maximum pixel value among the ${\sim}$6 Terapixels available and set the required number of bits.
In Table \ref{tab:max} we show the maximum pixel value from three representative GERLUMPH datasets, and in Table \ref{tab:bits} we show the largest integer number that can be represented by a fixed number of bits.
We can see that a 24-bit representation will be sufficient for the GERLUMPH maps, however, this representation does not correspond to any standard data type used by modern programming languages.
Therefore, the simplest solution is to choose the next available representation, 32-bits per pixel, which results in 381 Megabytes per map, and a total of 25 Terabytes for a collection of 70,000 maps.
If required, the total data size can be further reduced by a number of techniques described below.

\begin{table}
\begin{center}
\caption{The maximum pixel value and the maximum number of distinct pixel values among all maps of GD0, GD1 and GD2. The two values do not necessarily come from the same map of each dataset.}
\begin{tabular}{lccc}
       & max. ray-count   & max. distinct values \\
\hline
GD0    & 109117			  & 11776  \\
GD1    & 998878			  & 58042  \\
GD2    & 990652			  & 46306  \\
\hline
\end{tabular}
\label{tab:max}
\end{center}
\end{table}

\begin{table*}
\begin{center}
\caption{A different minimum number of bits is required according to the maximum integer number we want to represent in binary form. Given the values of Table \ref{tab:max}, this table shows how many bits have to be used to represent the pixel information in the GERLUMPH maps, either in ray-count or indexed format. The file size of a 10000$^2$-pixel map and the total size of a hypothetical collection of 70,000 such maps is shown as well.}
\begin{tabular}{lcccccc}
								& 14-bit & 16-bit   & 20-bit   & 22-bit	 & 24-bit   & 32-bit     \\
\hline
max. integer value				& 16383	 & 65535    & 1048575  & 4194304 & 16777215 & 4294967295 \\
ray-count						& no	 & no       & probably & yes     & yes      & yes        \\
indexed							& no 	 & probably & yes      & yes	 & yes      & yes        \\
data type available				& no     & yes      & no       & no      & no 	    & yes        \\
map size (Megabytes)			& 167	 & 191	    & 238      & 262	 & 286	    & 381	 \\
size of 70,000 maps (Terabytes)	& 11.2   & 12.8	    & 15.9     & 17.5	 & 19.1	    & 25.4	 \\
\hline
\end{tabular}
\label{tab:bits}
\end{center}
\end{table*}

First, we can make use of the fact that pixel values in a map do not cover continuously the range from the minimum to the maximum value.
For example, the maximum value in all of GD0 pixels is 109117, but the maximum number of distinct pixel values found in any GD0 map is only 11776 (see Table \ref{tab:max}).
Therefore, we can convert a map with ray-counts to an indexed map and an index file.
The index file will hold all the distinct ray-count values, each matched to an index running from zero to the maximum number of distinct ray-counts in the map.
The indexed map will have its pixel ray-count values replaced by the appropriate index.

We can then choose a binary representation for the indexed map with fewer bits than the ray-count map.
For example, the maximum index value in GD1 would be 58042, which can be represented by 16 bits, bringing the map size down to 191 Megabytes, and the total data size of a collection of 70,000 such maps to 13 Terabytes (see Tables \ref{tab:max} and \ref{tab:bits}).
The index file itself is small: if we use 16 bits to represent the index value and 32 bits for the corresponding ray-count value, we will end up with a file size of 0.34 Megabytes for the index file of a map with 58042 distinct pixel values (the maximum of GD1).
Using indexed maps could effectively halve the required disk space to store the GERLUMPH data.

\begin{figure}[t]
\begin{center}
\includegraphics[scale=0.18, angle=0]{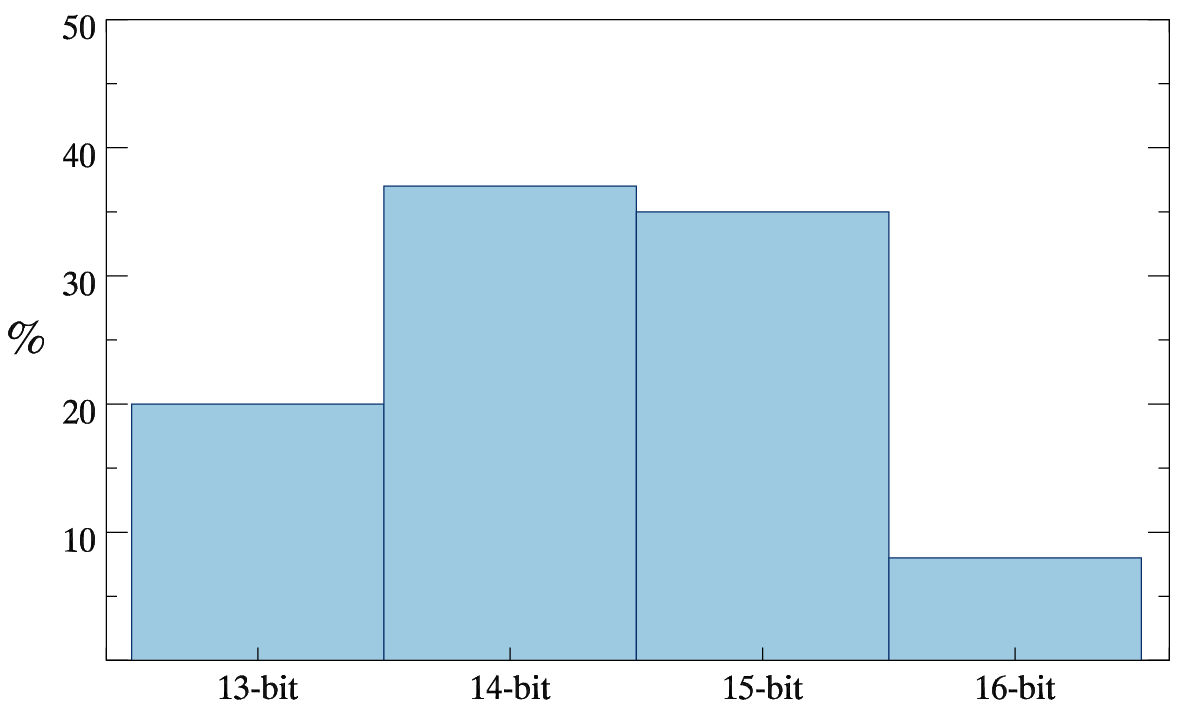}
\caption{Percentage of maps that can be represented by a given number of bits after being converted from ray-count to indexed format, according to the maximum number of distinct pixel values they contain. A sample of 50,000 existing GERLUMPH maps was used to produce this plot.}
\label{fig:hist}
\end{center}
\end{figure}

Finally, rather than using the standard 16-bit and 32-bit data types, we could define our own data types, using in each case the minimum required number of bits to represent a particular map (see Table \ref{tab:bits}).
In Figure \ref{fig:hist}, we can see the percentage of maps with the minimum number of bits required to represent them (in indexed format), from a sample of 50,000 existing GERLUMPH maps.
It is clear that in case we decide to use different number of bits per map, we will end up with a mix of different file sizes.

GERLUMPH map data are stored in 32-bit ray-count format, which is the simplest choice from a programming point of view.
Using indexed maps can reduce the file size to half, from 381 to 191 Megabytes per map.
We can use standard Unix tools to further compress the map binary files (such as {\tt gzip} or {\tt bzip2}), with 50 to 80 per cent compression ratios likely, as reported by \cite{Bate2012}.
To conclude, if required, it is possible to reduce the total data size of the GERLUMPH map data from 25 Terabytes to 2.6 - 6.4 Terabytes, without any loss of information.

\end{document}